\documentclass[notitlepage,letterpaper,aps,twocolumn,amsmath,amsfonts,nofootinbib,preprintnumbers,superscriptaddress,eqsecnum,secnumarabic]{revtex4-1}
\pdfoutput=1
\usepackage{natbib}
\usepackage{amssymb,amsmath,latexsym,mathrsfs}
\usepackage{bm}
\usepackage{url}
\usepackage{epsfig,graphicx,verbatim,xspace,multirow,mathtools}
\usepackage{array}
\usepackage{enumitem}
\usepackage{graphicx,slashed}
\usepackage[usenames,dvipsnames]{color}
\usepackage{todonotes}
\usepackage{color}
\usepackage{chngcntr}
\counterwithout{equation}{section} 
\usepackage[normalem]{ulem}
\usepackage[breaklinks,colorlinks,urlcolor=blue,citecolor=blue,linkcolor=blue]{hyperref}
\usepackage{epsfig,verbatim,xspace,multirow,mathtools}
\usepackage{subfiles}

\newcommand{\Eq}[1]{Eq.~\eqref{#1}}
\newcommand{\Fig}[1]{Fig.~\ref{#1}}

\newcommand{\ie}{\emph{i.e.~}}
\newcommand{\eg}{\emph{e.g.~}}

\begin{document}

\preprint{\tt KCL-2019-71}

\title{A CMB Search for the Neutrino Mass Mechanism \\ and its Relation to the Hubble Tension}

\author{Miguel Escudero}
\email{miguel.escudero@kcl.ac.uk}
\affiliation{King's College London, Department of Physics, Strand, London WC2R 2LS, UK}
\author{Samuel J. Witte}
\email{sam.witte@ific.uv.es}
\affiliation{Instituto de F\'{\i}sica Corpuscular (IFIC), CSIC-Universitat de Val\`encia, Spain}

\begin{abstract}
\noindent The majoron, a pseudo-Goldstone boson arising from the spontaneous breaking of global lepton number, is a generic feature of many models intended to explain the origin of the small neutrino masses. In this work, we investigate potential imprints in the Cosmic Microwave Background (CMB) arising from massive majorons, should they thermalize with neutrinos after Big Bang Nucleosynthesis via inverse neutrino decays. We show that {\tt Planck2018} measurements of the CMB are currently sensitive to neutrino-majoron couplings as small as $\lambda \sim 10^{-13}$, which if interpreted in the context of the type-I seesaw mechanism correspond to a lepton number symmetry breaking scale $v_L \sim \mathcal{O}(100) \, {\rm GeV}$. Additionally, we identify parameter space for which the majoron-neutrino interactions, collectively with an extra contribution to the effective number of relativistic species $N_{\rm eff}$, can ameliorate the outstanding $H_0$ tension. 
\end{abstract}

\maketitle


\noindent{\bf Introduction:}
Despite unambiguous evidence that at least two of the known neutrinos have a non-zero mass, the Standard Model (SM) is still lacking of an explanation of their origin. Perhaps more concerning, however, is the question of why neutrino masses are so much smaller than those of charged leptons. While many models have been proposed over the years to explain both the origin and smallness of the neutrino masses (see \eg~\cite{Mohapatra:1980yp,Babu:1988ki,Pilaftsis:1991ug,King:2003jb,Altarelli:2004za,Mohapatra:2005wg}), perhaps the most compelling class of models are those which invoke the so-called seesaw mechanism~\cite{Minkowski:1977sc,Mohapatra:1979ia,GellMann:1980vs,Yanagida:1980xy,Schechter:1980gr}. In such scenarios, the SM is augmented by heavy right-handed neutrinos carrying a Majorana mass term $m_N$, which naturally give rise to light neutrino masses $m_\nu$ of the order $\sim y_N^2 \, v_H^2 / m_N$, where $v_H \simeq 246$ GeV is the vacuum expectation value of the SM Higgs, and $y_N$ is the Dirac Yukawa coupling of the right-handed neutrinos. Generating the Majorana mass term necessary to implement the seesaw mechanism is often accomplished by introducing a new scalar that spontaneously breaks lepton number. Assuming that lepton number is a global symmetry, as in the SM, the spontaneous symmetry breaking (SSB) triggered by the scalar leads to the prediction of a pseudo-Goldstone boson, the so-called majoron~\cite{Chikashige:1980ui} (see also~\cite{Gelmini:1980re,Georgi:1981pg,Schechter:1981cv}).

The majoron is notoriously difficult to probe since it interacts very weakly with all SM particles, particularly with charged fermions $\lambda_{\phi e} \sim 10^{-20}$~\cite{Chikashige:1980ui}. However, measurements of the Cosmic Microwave Background (CMB) have reached a level of precision where small modifications to the neutrino sector may be discernible~\cite{Bashinsky:2003tk,Chacko:2003dt,Hannestad:2004qu,Hannestad:2005ex,Bell:2005dr,Friedland:2007vv,Brust:2017nmv,Anchordoqui:2012qu,Diamanti:2012tg,Archidiacono:2013dua,Oldengott:2014qra,Oldengott:2017fhy,Forastieri:2015paa,Lancaster:2017ksf,Kreisch:2019yzn,Park:2019ibn,Forastieri:2019cuf,Barenboim:2019tux,Escudero:2019gfk}. The effect of including majoron-neutrino interactions in the early Universe are twofold~\cite{Chacko:2003dt}: \textit{(i)} they lead to a non-standard expansion history after Big Bang Nucleosynthesis (BBN) and prior to recombination (generically amounting to $\Delta N_{\rm eff} \sim \mathcal{O}(0.1)$), and \textit{(ii)} they act to suppress the neutrino anisotropic stress energy tensor, and hence reduce neutrino free-streaming~\cite{Bashinsky:2003tk}. The idea of identifying features in the CMB arising from the majoron, and thus providing an indirect probe of the neutrino mass mechanism, was proposed at the start of the century~\cite{Chacko:2003dt}. However, until now, no rigorous cosmological implementation of this idea has been performed\footnote{Refs.~\cite{Bell:2005dr,Friedland:2007vv,Brust:2017nmv} explored the possibility that some component of radiation contained strong self-interactions;  this was accomplished by artificially setting to zero the multiples $\ell \ge 2$ in the Boltzmann hierarchy for the interacting radiation. This approach, however, cannot be applied (or mapped) into the scenario of~\cite{Chacko:2003dt}, since neutrino-majoron interactions rates are strongly time-dependent and not infinite in strength.}, nor has there been an analysis using real data.

Using {\tt Planck2018} data~\cite{Aghanim:2018eyx,Aghanim:2019ame}, we analyze a well-motivated region of parameter space in which majorons thermalize with neutrinos after BBN via inverse neutrino decay. We show that neutrino-majoron couplings as small as $10^{-13}$ can be robustly excluded with existing CMB data; future experiments, such as the Simons Observatory~\cite{Ade:2018sbj} and CMB-S4~\cite{Abazajian:2019eic}, which are aiming to probe the effective number of relativistic species $N_{\rm eff}$ at the sub-percent level, could have sensitivity to couplings as small as $10^{-14}$. If interpreted in the context of the type-I seesaw model, these couplings point toward a lepton number symmetry breaking scale of $\mathcal{O}(100) \, {\rm GeV}$ and $\mathcal{O}(1) \, {\rm TeV}$, respectively. Thus, quite remarkably, the CMB is providing an indirect probe of the neutrino mass mechanism at collider energy scales (albeit unaccessible to colliders due to their small couplings), but using feeble interactions with neutrinos in the early Universe. 

While the $\Lambda$CDM model has been incredibly successful at describing both high- and low-redshift cosmological observations, a concerning tension has recently emerged between the value of the Hubble constant $H_0$ inferred using early Universe observations (with data either from the CMB~\cite{Aghanim:2018eyx}, or by combining measurements from BBN with baryonic acoustic oscillations, \ie BAOs~\cite{Addison:2017fdm,Cuceu:2019for,Schoneberg:2019wmt}), and various local late Universe measurements performed using observations of type-Ia supernovae (see \eg \cite{Riess:2016jrr,Riess:2019cxk,Dhawan:2017ywl,Burns:2018ggj,Freedman:2019jwv,Yuan:2019npk}) and strong lensing~\cite{Bonvin:2016crt,Birrer:2018vtm,Rusu:2019xrq,Chen:2019ejq} (see \eg \cite{Verde:2019ivm} for an overview of the various measurements). The most prolific of these discrepancies is between the value inferred by {\tt Planck}, $H_0 = 67.4 \pm 0.5$ km/s/Mpc~\cite{Aghanim:2018eyx}, and that observed by SH$_0$ES collaboration, which relies on cepheids to calibrate the distance to type-Ia SN, who find a value of $H_0 = 74.0 \pm 1.4$ km/s/Mpc~\cite{Riess:2019cxk}. Depending both on the choice of distance calibration and how one chooses to combine datasets, the outstanding tension is determined to be at the level of $\sim 4 - 6\, \sigma$~\cite{Verde:2019ivm,Wong:2019kwg}. While it is of course possible that this tension is a consequence of unaccounted for systematics in either or both measurements, throughout this work we will take this discrepancy at face value and assume alternatively that this is an indication of new physics beyond the $\Lambda$CDM paradigm.

Various groups have attempted to resolve this issue by including additional contributions to $N_{\rm eff}$~\cite{Bernal:2016gxb,Mortsell:2018mfj,DEramo:2018vss,Escudero:2019gzq,Gelmini:2019deq}, strong neutrino self-interactions~\cite{Kreisch:2019yzn,Park:2019ibn}, hidden neutrino interactions~\cite{Archidiacono:2016kkh,DiValentino:2017oaw,Ghosh:2019tab}, exotic dark energy models~\cite{DiValentino:2016hlg,Qing-Guo:2016ykt,Karwal:2016vyq,Ko:2016uft,DiValentino:2017iww,DiValentino:2017rcr,Poulin:2018cxd,DiValentino:2019exe,Agrawal:2019lmo,Alexander:2019rsc,Lin:2019qug,Agrawal:2019dlm,DiValentino:2019ffd,Smith:2019ihp}, dark sector interactions~\cite{Bringmann:2018jpr,Pandey:2019plg,Raveri:2017jto,Yang:2019nhz}, and modified theories of gravity~\cite{Renk:2017rzu,Khosravi:2017hfi,Lin:2018nxe}. Most of these solutions are either incapable of resolving the tension fully~\cite{Martinelli:2019krf,Knox:2019rjx,Vagnozzi:2019ezj}, are experimentally constrained~\cite{Blinov:2019gcj}, are highly fine-tuned, or lack theoretical motivation. Perhaps the most simple, and thus theoretically appealing, solution which can ameliorate the $H_0$ tension to the level of $\sim 3\sigma$ is simply to postulate the existence of non-interacting dark radiation producing a shift in the radiation energy density relative to the value predicted in the Standard Model of $\Delta N_{\rm eff} \sim 0.25$. A more appealing, albeit far more problematic, solution was introduced in \cite{Kreisch:2019yzn}, where it was shown that strongly interacting 2-to-2 neutrino scatterings together with a contribution to $\Delta N_{\rm eff} \sim 1$ was able to fully resolve the tension; unfortunately, this solution requires neutrino couplings that are not phenomenologically viable~\cite{Blinov:2019gcj}, a value of $\Delta N_{\rm eff}$ excluded by BBN~\cite{Pitrou:2018cgg}, and is only successful at reducing the tension if CMB polarization data is neglected. Given that the majoron naturally contributes to $\Delta N_{\rm eff}$ at the level of $\sim 0.11$ via late-time thermalization and decay, and damps neutrino free-streaming in a manner similar to that of the strongly interacting neutrino solution, it is natural to ask whether 2-to-1 neutrino-majoron interactions are capable of further reducing the $H_0$ tension, beyond what is simply accomplished with $\Lambda$CDM + $\Delta N_{\rm eff}$. Indeed we show that including majoron-neutrino interactions broadens the posterior such that the $H_0$ tension can be further reduced, albeit only to the level of $2.5\,\sigma$, a level that is comparable with other viable solutions, such as early dark energy (see \eg\cite{Agrawal:2019lmo}).


~\\ \noindent{\bf Majoron Interactions:}
We parametrize the majoron-neutrino interaction as: 
\begin{align}
\mathcal{L} = i \,\frac{\lambda}{2} \, \phi \, \bar{\nu} \, \gamma_5 \, \nu\, ,
\end{align} 
where $\nu$ corresponds to a light neutrino mass eigenstate. The coupling $\lambda$, taken here to be universal, is typically intimately related to the mass of the active neutrinos $m_\nu$ and the scale at which lepton number is spontaneously broken, $v_L$. For example, in the type-I seesaw mechanism, $\lambda$ can be expressed as 
\begin{align}
\lambda = 2\, U^2 \, \frac{m_N}{v_L} \simeq 2 \, \frac{m_\nu}{v_L} \, ,
\end{align} 
where $U$ is the mixing between sterile and active neutrinos, and the last line follows from a condition in the type-I seesaw that $U^2 \sim m_\nu/m_N$~\cite{Mohapatra:2005wg}. Interestingly, for values of $v_L \sim v_H$ and neutrino masses consistent with current constraints, the value of $\lambda$ within this model can \eg naturally be of the order of $\lesssim 10^{-12}$, which happens to be around the region where inverse neutrino decays ($\bar{\nu}\nu \to \phi$) can thermalize light majorons after BBN, but prior to recombination. In what follows we will treat $\lambda$ as a free parameter to remain as model-independent as possible, and when appropriate, relate $v_L$ to $\lambda$ by considering the atmospheric mass splitting $m_\nu \sim \sqrt{|\Delta m_{\rm atm}^2|} \simeq 0.05\,\text{eV}$~\cite{pdg}. Namely, $v_L \simeq 1\,\text{TeV} \, (10^{-13}/\lambda)$.

\begin{figure*}
\centering
\includegraphics[width=0.95\textwidth]{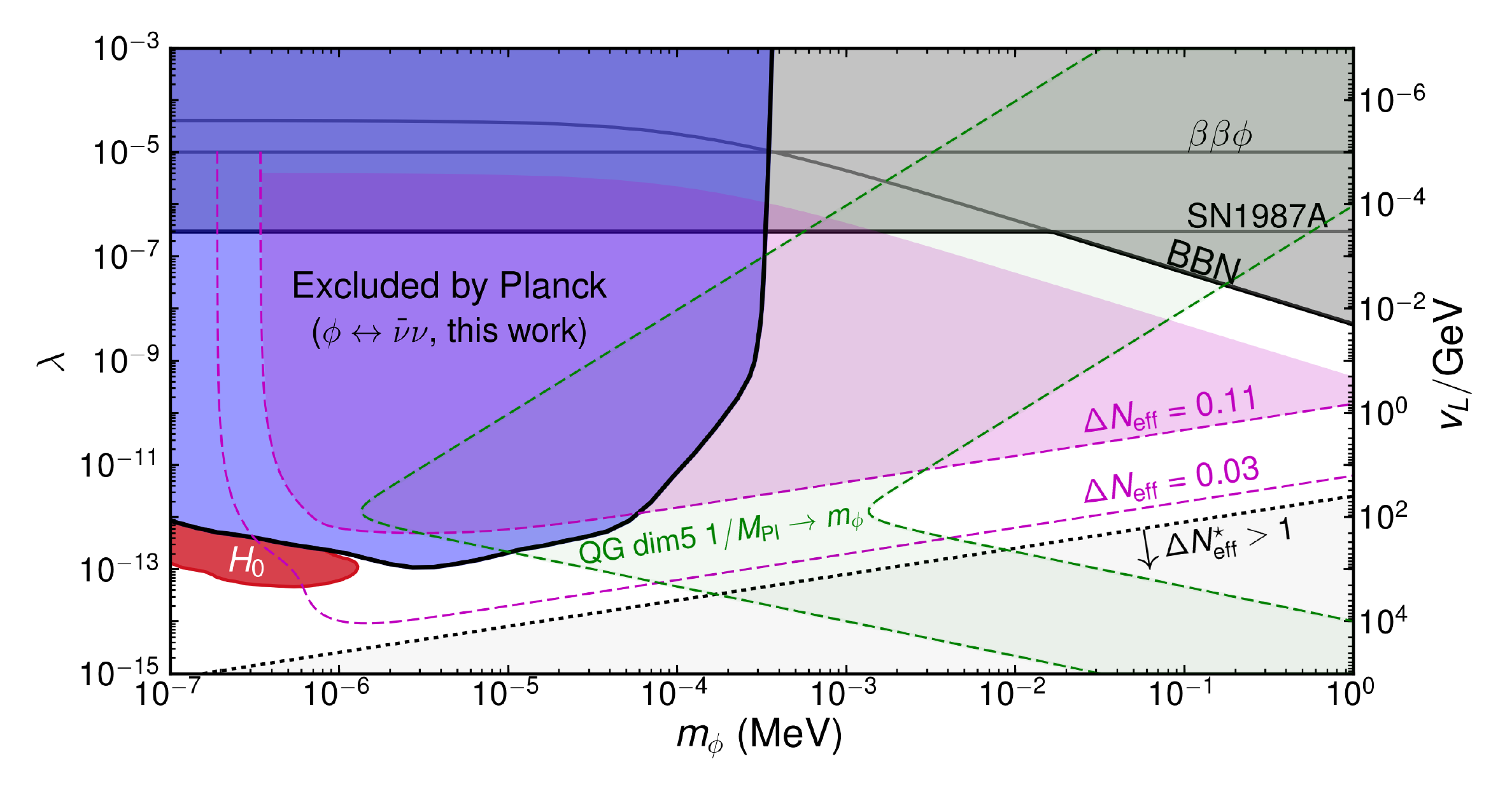} \vspace{-0.4cm}
\caption{Majoron parameter space. The left and right vertical axes correspond to the majoron-neutrino coupling and the scale at which lepton number is spontaneous broken in the type-I seesaw model respectively. Current constraints from KamLAND-Zen~\cite{Gando:2012pj}, BBN (see text), and SN1987A~\cite{Kachelriess:2000qc,Farzan:2002wx} are shown in grey. The pink region demarcates parameter space for which the majoron fully thermalizes after neutrino decoupling, leading to $\Delta N_{\rm eff} = 0.11$. The green band highlights the region of parameter space in which the majoron mass could arise from dim-5 Planck suppressed operators~\eqref{eq:mass_maj}. Shown in blue is the parameter space excluded in this work using {\tt Planck2018} data at 95\% CL. The parameter space below the black dotted line is excluded if there was a small but primordial population of thermal majorons. The region labeled `$H_0$' is the preferred $1\sigma$ contour for resolving the Hubble tension.}\label{fig:Majoron_easy}
\end{figure*}

~\\ \noindent{\bf The Majoron Mass:} Quantum gravity is expected to break all global symmetries~\cite{Banks:2010zn,Witten:2017hdv}, and hence the majoron should acquire a small but non-zero mass. Naively, one might expect the majoron mass to arise from dimension-five (dim-5) Planck scale suppressed operators~\cite{Rothstein:1992rh,Akhmedov:1992hi}. Should these dim-5 operators involve only the Higgs and the scalar responsible for the SSB of lepton number, the majoron mass is expected to be
\begin{align}\label{eq:mass_maj}
m_\phi^2 \sim {\frac{\beta}{0.1}} \, {\frac{v_H}{v_L}}\left[1  +35 \left(\frac{v_L}{v_H}\right)^{4} \right] \,\text{keV}^2 \, ,
\end{align}
where $\beta$ is the coupling constant of a given operator at the Planck scale -- which, for concreteness, we have assumed to be the same for all relevant dim-5 operators~\cite{Akhmedov:1992hi}. 
Of course, the actual details of the breaking of global symmetries by gravity depend upon the unknown quantum nature of the gravitational theory at the Planck scale~\cite{Kallosh:1995hi}; thus we treat $m_\phi$ as a free parameter in this work, centered approximately around the keV scale, but allowed to vary from $0.1 \, \text{eV}$ to $1\,\text{MeV}$.

\Fig{fig:Majoron_easy} contains a depiction of the majoron parameter space relevant for this work. In addition to highlighting parameter space currently excluded by SN1987A~\cite{Kachelriess:2000qc,Farzan:2002wx}, BBN (see Supplementary Material, and \eg~\cite{Chang:1993yp}), and KamLAND-Zen~\cite{Gando:2012pj}, we identify masses and couplings for which the majoron is consistent with arising from dim-5 Planck scale suppressed operators\footnote{This band is constructed by varying $\beta$ in \Eq{eq:mass_maj} between $10^{-6}$ and $1$, where the lower/upper limit has been chosen to be reflect the electron/top Yukawa coupling.}. We defer discussion of the remainder of this plot to later sections.

~\\ \noindent{\bf Model Extensions:} Looking forward, it may be interesting to consider the possibility that one of the active neutrinos is exactly massless, as this would decouple the lightest neutrino form the majoron, changing the cosmological evolution of the system. One could also conceive of the possibility of a multi-majoron system resulting from the SSB of a more complex flavor symmetry group in the neutrino sector~\cite{Chacko:2003dt}. In such a scenario, one could produce a more complicated thermalization history which produces step-like features in the evolution of the energy density, and damps the perturbations in a non-trivial manner. While these models are beyond the scope of the current work, they provide a clear extension of the ideas and prospects studied here.


~\\ \noindent{\bf Early Universe Cosmology:}
The collision terms governing the evolution of the neutrino and majoron phase space distributions are determined by the decay rate of the majoron into two neutrinos $\phi \to \bar{\nu} \nu$, given by
\begin{align}
\Gamma_\phi  = \frac{\lambda^2}{16\pi} m_\phi \, \sqrt{1-\frac{4m_\nu^2}{m_\phi^2}} \simeq \frac{\lambda^2}{16\pi} m_\phi \, ,
\end{align}
where in the last step we have considered $m_\nu \ll m_\phi$.
In order to model the time-dependent evolution of the number density and energy density of the system, we follow~\cite{Escudero:2020dfa} (see also~\cite{Escudero:2018mvt}) in assuming that all relevant species are characterized by a temperature $T_i$ and chemical potential $\mu_i$, and solve for their time evolution accounting for all relevant interactions\footnote{Ref.~\cite{Escudero:2020dfa} explicitly demonstrates that this method accurately reproduces a full numerical solution to the Liouville equation for the neutrino and majoron distribution functions within the relevant parameter space considered in this study.} (see Supplementary Material for details). If the majoron is sufficiently heavy and interactions sufficiently strong, the majorons may begin to thermalize prior to or during BBN, leading to an enhanced expansion history of the Universe that would alter the formation of the light elements. For small couplings and masses ($\lambda \lesssim 10^{-5}$ and $\lambda \lesssim 10^{-10} \,\text{MeV}/m_\phi$), majorons thermalize with neutrinos after BBN, and when the majorons become non-relativistic at $T_\nu \sim m_\phi/3$, they decay out of equilibrium to neutrinos leading to a small enhancement in $N_{\rm eff}$, which asymptotes to $\Delta N_{\rm eff}= 0.11$. We identify in \Fig{fig:Majoron_easy} a shaded pink region for which full thermalization is achieved after BBN. For yet smaller couplings, partial thermalization can be achieved; the dashed pink line in \Fig{fig:Majoron_easy} identifies majorons that never thermalize, but augment $N_{\rm eff}$ to a level that may be observable with CMB-S4 experiments~\cite{Abazajian:2016yjj}.

We model the phase space perturbations by considering the coupled neutrino-majoron fluid, and approximate the entire system as being massless\footnote{The error introduced by neglecting neutrino masses is the Boltzmann hierarchy is expected to be entirely negligible given current constraints on $\sum m_\nu <  0.12$ eV~\cite{Aghanim:2018eyx}, see also~\cite{Vagnozzi:2017ovm,Loureiro:2018pdz,RoyChoudhury:2019hls,Vagnozzi:2019utt}.}. Despite the fact that the temperature of the Universe eventually becomes similar to the majoron mass, the majoron contribution to the energy density of the neutrino-majoron system is never larger than 10$\%$. We have explicitly verified that the equation of state $\omega = (p_\phi + p_\nu) / (\rho_\phi + \rho_\nu)$ and the speed of sound $c_s^2 = \delta(p_\phi + p_\nu) / \delta(\rho_\phi + \rho_\nu)$ deviate by less than $3\%$ with respect to that of an ultra-relativistic fluid, \ie $\omega = c_s^2 = 1/3$ (see Supplementary Material). Additionally, we adopt the relaxation time approximation for the collision term~\cite{Hannestad:2000gt}, which has been shown to accurately reproduce the full solution in similar scenarios~\cite{Oldengott:2014qra,Oldengott:2017fhy}. The above simplifications allow us to express the density contrast $\delta$, the fluid velocity $\theta$, the shear $\sigma$, and the higher anisotropic moments in the synchronous gauge as~\cite{Ma:1995ey,Hannestad:2000gt}:
\begin{subequations}\label{eq:Hierarchy}
\begin{align}
\dot{\delta}_{\nu\phi} &= - \frac{4}{3}\theta_{\nu\phi} - \frac{2}{3}\dot{h} \,, \\
\dot{\theta}_{\nu\phi} &= k^2 \left(\frac{1}{4}\delta_{\nu\phi} -\sigma_{\nu\phi} \right)  \,,  \\
\begin{split}
\dot{F}_{\nu\phi} {}_{2} &= 2\dot{\sigma}_{\nu\phi} = \frac{8}{15}\theta_{\nu\phi}-\frac{3}{5}kF_{\nu\phi \,3} \\ & +\frac{4}{15}\dot{h}+\frac{8}{5}\dot{\eta}  - 2\, a\, \Gamma {\sigma}_{\nu\phi}   \,, \end{split} \\
\begin{split}
\dot{F}_{\nu\phi\,\ell} &= \frac{k}{2\ell + 1} \left[ \ell \, {F}_{\nu\phi \,(\ell-1)} - (\ell +1){F}_{\nu\phi \,(\ell+1)}   \right] \\ & - a \,  \Gamma \, {F}_{\nu\phi\,\ell}  \, \hspace{.6cm} {\rm for} \hspace{.3cm}  \ell \geq 3 \, . \end{split}
\end{align}
\end{subequations}
Here, $h$ and $\eta$ account for the metric perturbations, $k$ is a given Fourier mode, ${F}_{\nu\phi\,\ell}$ represents the $\ell^{\rm th}$ multipole, $a$ the scale factor, and $\Gamma$ is the interaction rate accounting for inverse neutrino decays and majoron decays, given by
\begin{align}
\Gamma   = \frac{\Gamma_\phi}{2} \, \frac{m_\phi^2}{T_\nu^2} \, e^{\frac{\mu_\nu}{T_\nu}} \,K_1\left(\frac{m_\phi}{T_\nu}\right) \, ,
\end{align}
where $K_1$ is the modified Bessel function of the first kind. For convenience one can approximate $e^{\frac{\mu_\nu}{T_\nu}}  \simeq 1$, and $T_{\gamma}/T_\nu \simeq 1.4$ -- we have verified that this introduces a negligible error in the final result. In Eqns.~\eqref{eq:Hierarchy} all derivatives are understood to be with respect to conformal time.


~\\ \noindent{\bf Analysis:}
In order to efficiently scan the parameter space of interest, we define an effective interaction $\Gamma_{\rm eff}$ in terms of the majoron mass and coupling as
\begin{align}\label{eq:gamma_eff2}
\Gamma_{\rm eff} = \left(\frac{\lambda}{4\times 10^{-12}} \right)^2 \, \left(\frac{1 \, {\rm keV}}{m_\phi} \right) \, .
\end{align}
This effective interaction is defined such that for $\Gamma_{\rm eff} \gtrsim 1$ majorons thermalize in the early Universe. We perform runs with two distinct sets of priors: the first is used to place constraints on majoron models producing strong modifications to the neutrino perturbations, and the second is used to identify parameter space for which the $H_0$ tension can be ameliorated. For both sets of runs, we adopt log-flat priors in $\lambda$ or $\Gamma_{\rm eff}$ and $m_\phi$ spanning
\begin{subequations}
\begin{align}
\log_{10}\left(\lambda \right) &= [ -15,-6]\, ,\\
\log_{10}\left(m_\phi/\text{eV}\right) &= [ -2,3]  \, ,
\end{align}
\end{subequations}
and
\begin{subequations}
\begin{align}
\log_{10}\left(\Gamma_{\rm eff}\right) &= [ -4,4] \, ,\\
\log_{10}\left(m_\phi/\text{eV}\right) &= [ -2,2]  \, ,
\end{align}
\end{subequations}
respectively. In addition to these two parameters, we also allow for the possibility of extra relativistic and non-interacting degrees of freedom. We allow $\Delta N_{\rm eff}$ to vary linearly between $-2 \leq \Delta N_{\rm eff} \leq 4$, and treat this additional radiation as free streaming. This additional contribution to $N_{\rm eff}$ should not be considered {\emph{ad hoc}}, but rather a natural expectation of majoron models. For example, should the reheating temperature be above the mass of right handed neutrinos, a thermal population of majorons produced in the early Universe may come to dominate the energy density of the Universe, producing nearly arbitrarily large contributions to $\Delta N_{\rm eff}$. Such an effect becomes increasingly important for feeble interactions, such that an effective lower bound can be placed on the the neutrino-majoron interaction -- needless to say, however, this bound is inherently dependent on pre-BBN cosmology. We include in \Fig{fig:Majoron_easy} a line, labeled $\Delta N_{\rm eff}^*$, that identifies parameter space for which the contribution to $\Delta N_{\rm eff}$ from a primordial population of majorons would be excluded by {\tt Planck} and measurements of large scale structure. We include a more comprehensive discussion of this effect in the Supplementary Material.

\begin{figure}
\centering
\includegraphics[width=0.48\textwidth]{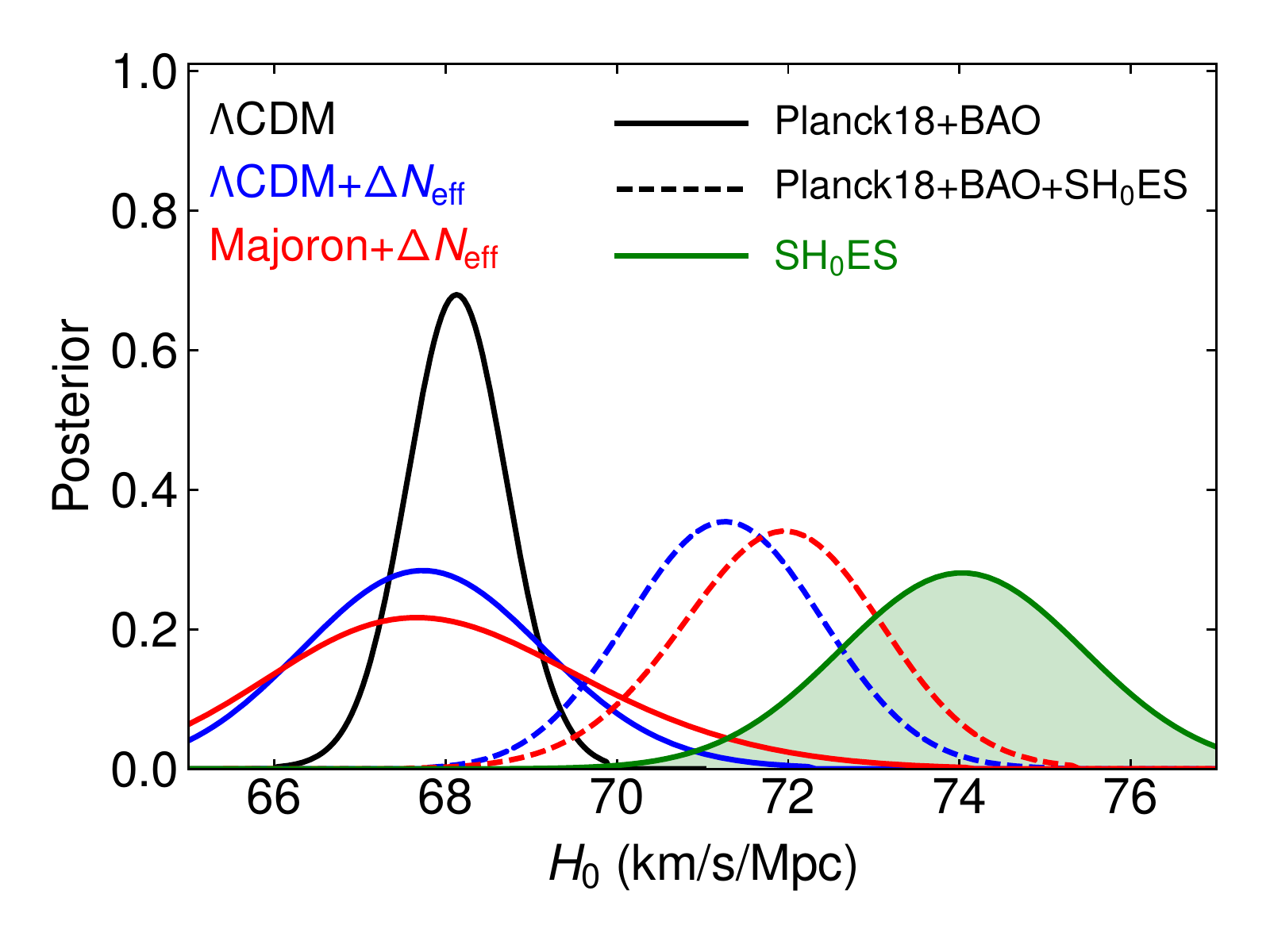}\vspace{-0.3cm}
\caption{$H_0$ posteriors for $\Lambda$CDM (black), $\Lambda$CDM + $\Delta N_{\rm eff}$ (blue), and majoron + $\Delta N_{\rm eff}$ (red), using {\tt Planck2018} + BAO (solid) and including a gaussian likelihood for SH$_0$ES (dashed). SH$_0$ES posterior shown for comparison in green. See Table~\ref{table:param_values} for best-fit values and $1\sigma$ uncertainties. The red solid line roughly corresponds to $H_0 = 68.0 \pm 1.9$ km/s/Mpc and hence is in $2.5\sigma$ tension with the SH$_0$ES measurement.}\label{fig:H0post}
\end{figure}

\vspace{0.4cm}
~\\ \noindent{\bf Results and Conclusions:}
After implementing the above modifications to both the energy density and neutrino-majoron perturbations in {\tt CLASS}~\cite{Blas:2011rf,Lesgourgues:2011re}, we perform an MCMC with {\tt Montepython}~\cite{Brinckmann:2018cvx,Audren:2012wb} using the {\tt Planck-2018} TTTEEE+lowlTT+lowE+lensing likelihood~\cite{Aghanim:2019ame}, including data on BAOs from the 6DF galaxy survey~\cite{Beutler:2011hx}, the MGS galaxy sample of SDSS~\cite{Ross:2014qpa}, and from the CMASS and LOWZ galaxy samples of BOSS DR12~\cite{Alam:2016hwk}, both including and excluding a Gaussian contribution to the likelihood on $H_0$ from SH$_0$ES~\cite{Riess:2019cxk}, taken to have a mean value and standard deviation of $74.0$ and $1.4$ km/s/Mpc. All MCMCs have been run until the largest Gelman-Rubin coefficient was $R-1 < 0.03$ or better. In  Table~\ref{table:param_values} we outline all relevant cosmological parameters for the analyses of Planck 2018+BAO+SH$_0$ES data.

In \Fig{fig:Majoron_easy} we show the $95\%$ exclusion contours derived in this work, and the 1$\sigma$ contour for parameter space preferred from the fit including the SH$_0$ES likelihood. We derive the $95\%$ CL exclusion contours using only {\tt Planck} data in order to remain conservative, and note that including \eg BAO data leads to a minor strengthening of this contour. Interestingly, the results obtained here illustrate that {\tt Planck} has already begun to significantly probe well-motivated regions of parameter space in which the majoron mass arises from $\text{dim-5}$ Planck scale suppressed operators. If interpreted in terms of the type-I seesaw model, current CMB observations are now probing lepton symmetry breaking scales $\mathcal{O}(100)$ GeV, with future CMB experiments potentially reaching the level of $\sim 10$ TeV. Before continuing, we would like to emphasize that the constraints derived in this work are both stringent and robust over wide regions of parameter space. For example, a majoron of $m_\phi = 1\,\text{eV}$ and $\lambda = 10^{-11}$ is excluded by more than $5\sigma$.

In \Fig{fig:H0post} we show the posterior distributions for $\Lambda$CDM, $\Lambda$CDM with a floating value of $\Delta N_{\rm eff}$, and the majoron + $\Delta N_{\rm eff}$, each including and excluding the SH$_0$ES likelihood. The SH$_0$ES posterior is shown for comparison. Including the majoron broadens the posterior and induces a minimal shift of the central value to large $H_0$, an effect which is more visible when the SH$_0$ES likelihood is included. While the difference induced by including the majoron is not enormous, the $H_0$ tension can be reduced from $4.4\,\sigma$ to $2.5\,\sigma$ when neutrino-majoron interactions, and an additional contribution to dark radiation, are included. 

By performing a MCMC including the SH$_0$ES likelihood, we find that a scenario with $\Delta N_{\rm eff} = 0.52\pm 0.19$, $0.1\,\text{eV} <m_\phi < 1\,\text{eV}$, and coupling strengths $\lambda \sim (10^{-14}-10^{-13} ) \, (\text{eV}/m_\phi)$ -- as highlighted in red in Fig.~\ref{fig:Majoron_easy} -- would render a posterior for $H_0$ of $71.9\pm1.2\,\text{km/s/Mpc}$ and an overall improvement of $\Delta \chi^2 \simeq -12.2$ with respect to $\Lambda$CDM. We remind the reader here that, because of the residual 2.5$\sigma$ tension, it may not be entirely meaningful to combine the partially discrepant datasets, and thus care should be given in the interpretation of this region. Notice that the improvement in the $\chi^2$ does not \emph{exclusively} arise from the shift in $H_0$; this can be seen from the fact that the contribution of the CMB likelihood in the Majoron+$\Delta N_{\rm eff}$ is less than that of $\Lambda$CDM. Interestingly, this region of parameter space corresponds to lepton number symmetry breaking scales in the type-I seesaw near the electroweak scale. Furthermore, it is worth emphasizing that unlike the strongly interacting neutrino solution proposed in~\cite{Kreisch:2019yzn} (defined by a 2-to-2 neutrino contact interaction), the solution proposed here is robust to the inclusion of polarization data, is phenomenologically viable, and is theoretically motivated.

An important comment on the consistency of this type of solution is necessary. If the contribution to $N_{\rm eff}$ is of primordial origin, then successful BBN excludes values of $\Delta N_{\rm eff} \gtrsim 0.4$ at $T \sim {\rm MeV}$~\cite{Pitrou:2018cgg,Cyburt:2015mya,Berlin:2019pbq}. In addition, including a floating value of $\Delta N_{\rm eff}$ in the CMB analysis can induce a shift in the preferred value of $\Omega_bh^2$, which is also constrained by BBN. In the Supplementary Material, we address the extent to which the parameter space in the $\Delta N_{\rm eff}-\Omega_bh^2$ plane preferred by the CMB fit is compatible with expectations of BBN. 

Evidence for the existence of the majoron, arising from the spontaneous breaking of global lepton number, would provide a strong clue to the origin of the neutrino masses. In this work we have looked at the extent to which CMB measurements have probed the existence of such a particle through its impact on the expansion history of the Universe and its interactions with neutrinos. We show that there exists a broad range of well-motivated parameter space that is now excluded using {\tt Planck2018} measurements of the CMB power spectrum. Furthermore, we identify a region in which the majoron interactions help ameliorate the outstanding $H_0$ tension to a level that is beyond what is simply accomplished by including $\Delta N_{\rm eff}$. If confirmed, the $H_0$ tension could be providing the first insight into the origin of the small neutrino masses.

\begin{table*}[t]
  \begin{tabular}{l|c|c|c}
    \hline\hline
    Parameter            				 & $\Lambda$CDM 				& $\Lambda$CDM+$\Delta N_{\rm eff}$     	 &Majoron + $\Delta N_{\rm eff}$ \\ \hline
    $\Delta N_{\rm eff} $      		&$-$ 						& $0.43~(0.358)\pm0.18$           		  & $0.52~(0.545)\pm0.19$ \\
     $m_\phi/\text{eV}  $  			    &$-$     						& $-$                                        			  & $ (0.33) $  \\
    $\Gamma_{\rm eff} $    			& $-$ 						&$-$                                         			&  $(8.1)$\\
    $100\,\Omega_b h^2$  			&$2.252~(2.2563)\pm0.016$ 		& $2.270~(2.2676)\pm0.017$ 			& $2.280~(2.2765)\pm0.02$ \\
    $\Omega_{\rm cdm} h^2$		&$0.1176~(0.11769)\pm0.0012$ 	& $0.125~(0.1243)\pm0.003$                         & $0.127~(0.1279)\pm0.004$\\
    $100~\theta_s$         			& $\,\,\,1.0421~(1.04223)\pm0.0003\,\,\,$ 	& $\,\,\,1.0411~(1.04125)\pm0.0005\,\,\,$ 		    & $\,\,\,1.0410~(1.04102)\pm0.0005\,\,\,$\\
    $\ln(10^{10}A_{s }) $   			 & $3.09~(3.1102)\pm0.03$ 		&$3.10~(3.072)\pm0.03$ 				    & $3.11~(3.116)\pm0.03$ \\
    $n_s$               				 & $0.971~(0.9690)\pm0.004$ 		&$0.981~(0.9780)\pm0.006$ 			   & $0.990~(0.99354)\pm0.010$ \\
    $\tau_{\rm reio}$         			& $0.051~(0.0500)\pm 0.008$    	 & $0.052~(0.0537)\pm 0.008$ 		  	&$0.052~(0.0576)\pm0.008$\\
    \hline
    $H_0$                 				 &$68.98~(69.04)\pm0.57$ 			&$71.27~(70.60)\pm1.1$ 			                 & $71.92~(71.53)\pm1.2$ \\
    \hline
    $(R-1)_{\rm min}$  			&      	  0.009						&0.009                                                              		&  0.03    		       \\
    $ \chi^2_{\rm min} $ high-$\ell$  	&      2341.56   						&     2345.39                                                       		&     2338.84  		       \\
    $ \chi^2_{\rm min} $ lowl	&         22.45						&        21.56                                                      		&       20.81		       \\
    $ \chi^2_{\rm min} $ lowE  	&   395.72      						&          395.89                                                		&      396.40 		       \\
    $ \chi^2_{\rm min} $ lensing  	&       9.91 						&        9.21                                                     		&    10.69   		       \\
    $ \chi^2_{\rm min} $ BAO  	&         4.74						&       4.5                                                        		&   4.69   		       \\
    $ \chi^2_{\rm min} $ SH$_0$ES  	&         12.34						&        5.82                                                        			&       3.10		       \\
        $ \chi^2_{\rm min} $ CMB  &         2769.6 					&        2772.1                                                        		&      2766.7		       \\
        $ \chi^2_{\rm min} $ TOT  	&         2786.7						&        2782.4                                                      		&      2774.5		       \\\hline
$\chi^2_{\rm min}-\chi^2_{\rm min}|^{\Lambda {\rm CDM}}$  		&  	$0$					                                       	&-4.3  		  &  -12.2                     		   \\
    \hline\hline

  \end{tabular}
  \caption{Mean (best-fit) values with $\pm 1\sigma$ errors of the cosmological 
parameters reconstructed from our combined analysis of 
Planck2018+BAO+SH$_0$ES data in each scenario. For comparison, the best-fit $\chi^2$ we find for $\Lambda$CDM using Planck2018+BAO data only with $(R-1)_{\rm min} = 0.007$ is: $\chi^2_{{\rm high}-\ell} = 2340.25$, $\chi^2_{{\rm lowl}} = 22.54$, $\chi^2_{{\rm lowE}} = 395.74$, $\chi^2_{{\rm lensing}} = 8.92$, $\chi^2_{{\rm BAO}} = 3.57$, $ \chi^2_{\rm CMB} = 2767.45 $.  }
  \label{table:param_values}
\end{table*}

\section*{Acknowledgments}
The authors thank Isabel Oldengott and Olga Mena for useful discussions. ME is supported by the European Research Council under the European Union's Horizon 2020 program (ERC Grant Agreement No 648680 DARKHORIZONS). SJW would like to thank ME and the TPPC group at King's College, as well as the Fermilab theory group, for the hospitality during the extended stays that lead to the completion of this work. SJW acknowledges support under Spanish grants FPA2014-57816-P and FPA2017-85985-P of the MINECO and PROMETEO II/2014/050 of the Generalitat Valenciana, and from the European Union's Horizon 2020 research and innovation program under the Marie Sk\l{}odowska-Curie grant agreements No.\ 690575 and 674896. 

\bibliographystyle{JHEP}
\bibliography{biblio}

\maketitle
\onecolumngrid
\newpage
\begin{center}
\vspace{0.05in}
{ \it \large Supplementary Material}\\ 
\vspace{0.05in}
{Miguel Escudero and Samuel J. Witte}
\end{center}
\onecolumngrid
\setcounter{equation}{0}
\setcounter{figure}{0}
\setcounter{section}{0}
\setcounter{table}{0}
\makeatletter
\renewcommand{\theequation}{S\arabic{equation}}
\renewcommand{\thefigure}{S\arabic{figure}}
\renewcommand{\thetable}{S\arabic{table}}

The Supplementary Material section contains additional information justifying various comments and statements asserted in the text, and outlining various computational details relevant for the reproducibility of this work. We also expand briefly on various phenomenological aspects. We begin by providing details on the computation of the background evolution and CMB phenomenology. We then discuss the derivation of the BBN constraint shown in \Fig{fig:Majoron_easy}, and finally discuss the implications of a primordial majoron population produced in the early Universe, which can be relevant should the reheating temperature of the Universe be sufficiently high.


~\\ \noindent{\bf Background Evolution:}
We follow~\cite{Escudero:2020dfa} (see also~\cite{Escudero:2018mvt}) and assume throughout that the distribution function for all relevant species can be characterized by their temperature $T_i$ and chemical potential $\mu_i$. The time evolution equations for each of such quantities reads~\cite{Escudero:2020dfa}:
\begin{align}
\frac{dT}{dt} &= \frac{1}{(\partial_\mu n) \, (\partial_T \rho) - (\partial_T n) \, (\partial_\mu \rho)} \left[ -3 \,H \, \left((p+\rho) \partial_\mu n - n \, \partial_\mu \rho \right)+ (\partial_\mu n) \, (\partial_t \rho)   - (\partial_\mu \rho) \, (\partial_t n)  \right]
 \, ,\label{eq:dT_dt_simple} \\
 \frac{d\mu}{dt} &= \frac{-1}{(\partial_\mu n) \, (\partial_T \rho) - (\partial_T n) \, (\partial_\mu \rho)}  \left[ -3 \,H \, \left((p +\rho) \, \partial_T n - n \, \partial_T \rho  \right) + (\partial_T n) \, (\partial_t \rho)  - (\partial_T \rho) \, (\partial_t n) \right]   \, , \label{eq:dmu_dt_simple} 
\end{align}
where $n,\,\rho,$ and $p$ correspond to the number, energy and pressure density of the given species, $H$ is the Hubble parameter, and $\delta_t \rho$ and $\delta_t n$ are the energy and number density exchange rates. Here the chemical potentials are the same for neutrinos and antineutrinos since they are produced at the same rates. Since we are exclusively interested in $1\leftrightarrow 2$ processes, within the Maxwell-Boltzmann approximation, we can express the energy and number density exchange rates as~\cite{Escudero:2020dfa}:
\begin{align}
\delta_t n  &= 3\frac{\Gamma_\phi   m_\phi^2 }{2 \pi ^2}\left[T_\nu e^{\frac{2 \mu_\nu }{T_\nu}} K_1\left(\frac{m_\phi}{T_\nu}\right)-T_\phi e^{\frac{\mu_\phi}{T_\phi}} K_1\left(\frac{m_\phi}{T_\phi}\right)\right]\,,\\ 
\delta_t \rho &= 3\frac{\Gamma_\phi  m_\phi^3 }{2 \pi ^2}\left[T_\nu e^{\frac{2 \mu_\nu }{T_\nu}} K_2\left(\frac{m_\phi}{T_\nu}\right)-T_\phi e^{\frac{\mu_\phi}{T_\phi}} K_2\left(\frac{m_\phi}{T_\phi}\right)\right]\,.
\end{align}
Conservation of energy and number of particles in the $\phi \to \bar{\nu}\nu$ process implies
\begin{align}
\delta_t n_\nu = - 2 \, \delta_t n_\phi  \,, \qquad \delta_t \rho_\nu = - \delta_t \rho_\phi\,.
\end{align}

The above system of equations are solved\footnote{Solver publicly available with the~\href{https://github.com/MiguelEA/nudec_BSM}{NUDEC\_BSM} code~\cite{Escudero:2018mvt,Escudero:2020dfa}.} starting from a sufficiently large temperature such that the majoron population is negligible in the plasma, and with initial conditions obtained from neutrino decoupling within the SM~\cite{Escudero:2020dfa}:
\begin{align}
T_\gamma/T_\nu = 1.39440\,, \qquad T_\nu/\mu_\nu = -239.4\,, \qquad T_\phi/T_\gamma = 10^{-5} \,, \qquad \mu_\phi/T_\gamma = -10^{-3}\,.
\end{align}
This system is evolved until the maximum time between $T_\gamma = m_\phi/20$ and $t = 20\times \tau_\phi$ to ensure that the majoron population disappeared from the Universe. We have ensured that the continuity equation $d \rho_{\rm tot}/dt = -H(\rho_{\rm tot} + p_{\rm tot})$ is fulfilled at each integration time step with a relative accuracy of $10^{-5}$ or better.  Should the majoron thermalize with the neutrinos while relativistic, occurring for $\tau_\phi \ll 1/H(T=m_\phi/3)$, one can solve for the resulting temperature and chemical potential of the joint neutrino-majoron system. Imposing conservation of energy and number density:
\begin{align}
\rho_\nu(T_\nu, 0) &= \rho_\nu(T_{\rm eq}, \mu_{\rm eq}) + \rho_\phi(T_{\rm eq}, 2\mu_{\rm eq})\,, \\
n_\nu(T_\nu, 0) &= n_\nu(T_{\rm eq}, \mu_{\rm eq}) + 2 n_\phi(T_{\rm eq}, 2\mu_{\rm eq}) \, ,
\end{align}
one finds that the equilibrium temperature and chemical potential are given by
 \begin{align}
T_{\rm eq} = 1.1204 \, T_\nu, \qquad \mu_{\rm eq} = -0.6445 \, T_\nu \, .
\end{align}
These values in turn imply that the maximum energy density contained in the majoron species is 
\begin{align}\label{eq:En_max}
\rho_\phi(T_{\rm eq}, \mu_{\rm eq}) \simeq 0.09Â \times \rho_\nu (T_{\rm eq}, \mu_{\rm eq}) < 0.045 \times \rho_{\rm tot}\,.
\end{align}
In the case in which the majoron thermalizes with the neutrinos (\ie $\tau_\phi < 1/H(T=m_\phi/3)$ equivalently to $\Gamma_{\rm eff} > 1$) and then decays, one finds the following asymptotic values for the temperature and chemical potential: 
\begin{align}\label{eq:asymp_t_mu}
T_\gamma/T_\nu = 1.33457\,, \qquad T_\nu/\mu_\nu = -6.728\, . 
\end{align}
\Eq{eq:asymp_t_mu} can then be used to compute $\Delta N_{\rm eff}$ at the time of recombination and the energy density stored in neutrinos today ($\Omega_\nu h^2$), the values of which are given by:
\begin{align}
 \Delta N_{\rm eff} = 0.11 \,, \qquad \Omega_{\nu} h^2 = \frac{\sum_\nu m_\nu}{93.2 \,\text{eV}}\,.
 \end{align}

\begin{figure}
\centering
\begin{tabular}{cc}
\hspace{-0.8cm} \includegraphics[width=0.53\textwidth]{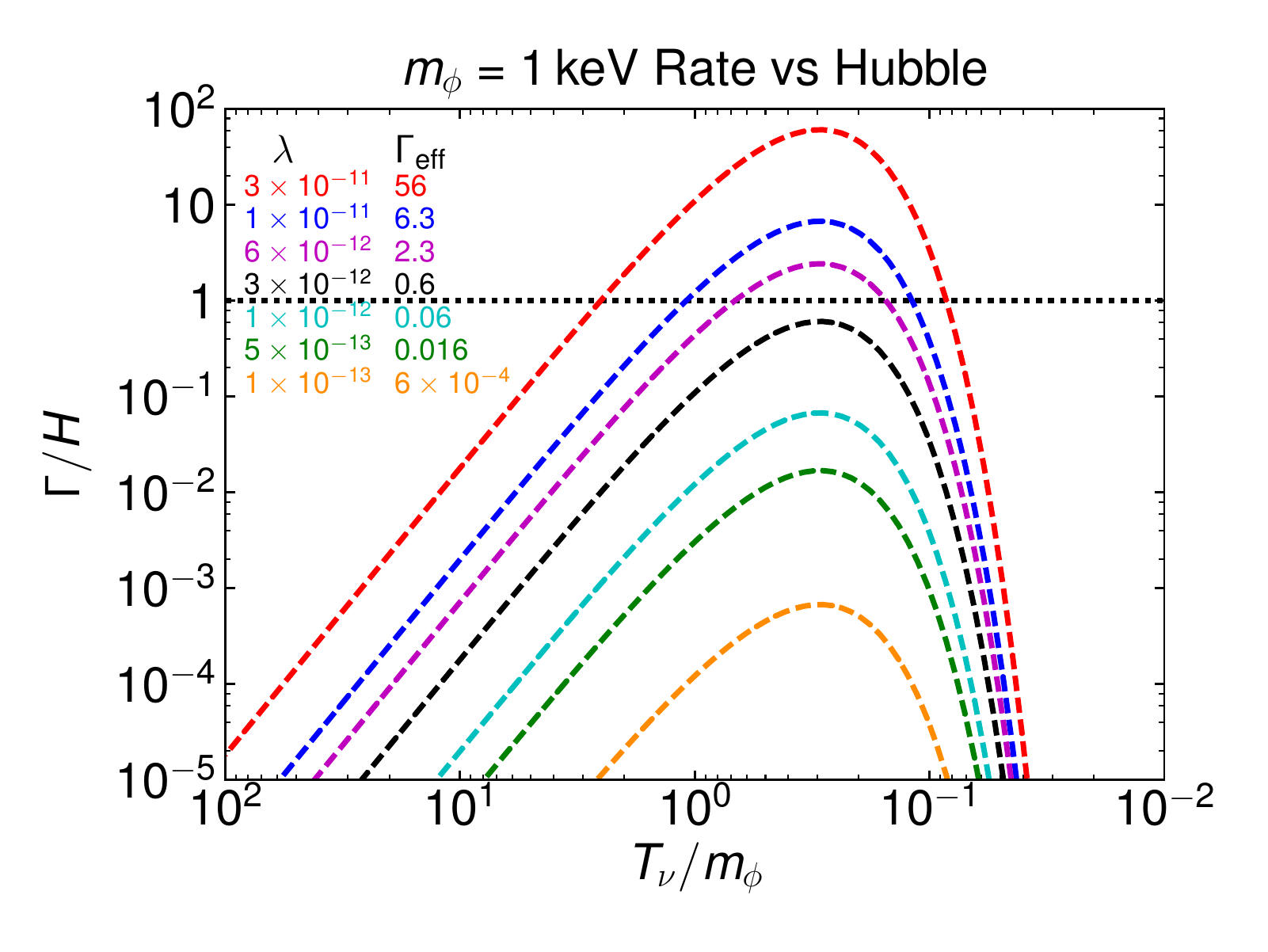} & \hspace{-0.8cm} \includegraphics[width=0.53\textwidth]{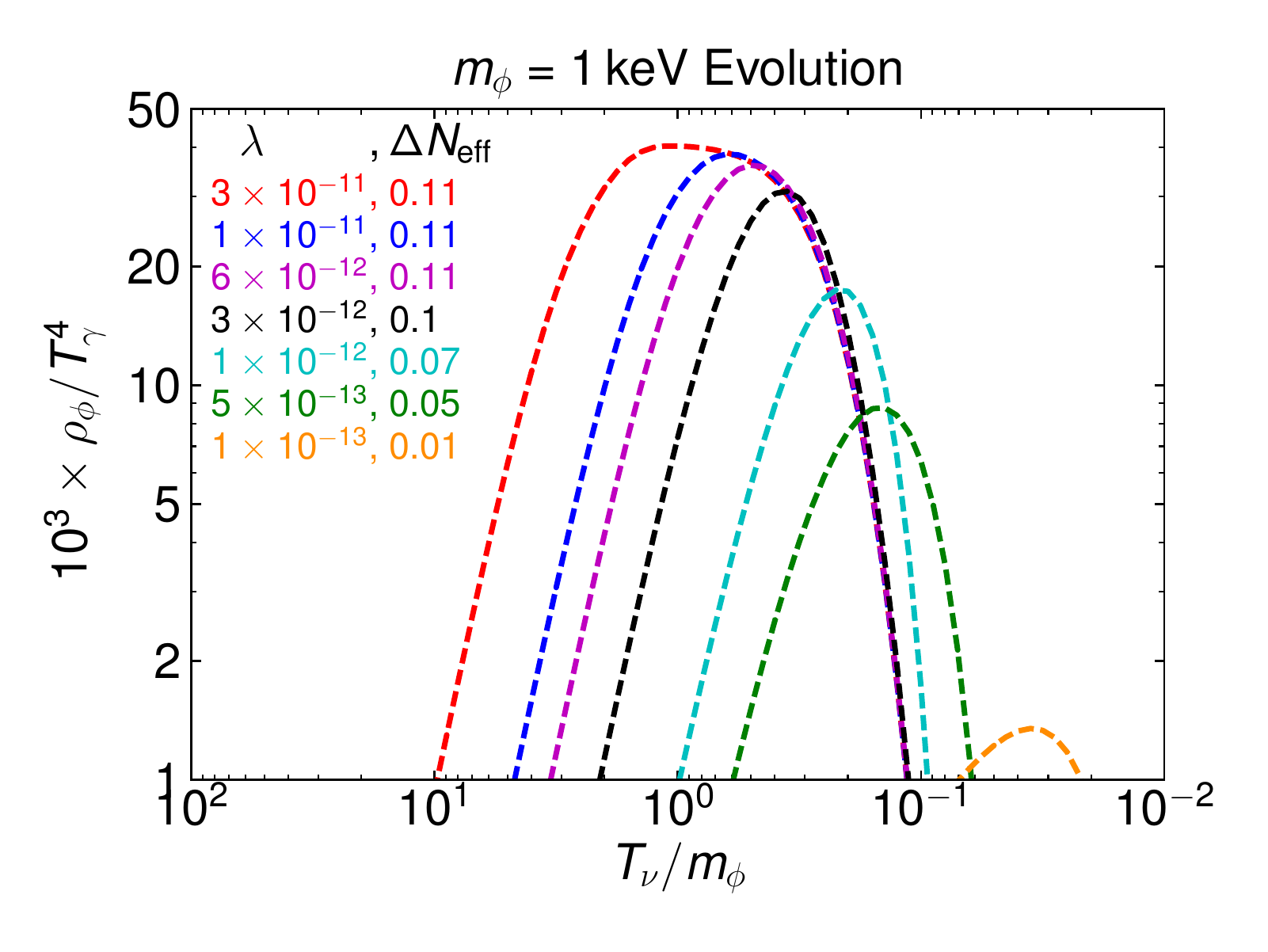}
\end{tabular}\vspace{-0.5cm}
\caption{\label{fig:Majoron_evolution} \textit{Left panel:} Ratio of majoron-neutrino interaction rate to Hubble expansion rate. \textit{Right panel:} Majoron energy density evolution as a function of the interaction strength. }
\end{figure}

In \Fig{fig:Majoron_evolution} we compare the interaction rate of a $1$ keV majoron to the Hubble expansion rate for various values of $\lambda$. When $\Gamma/H \gtrsim 1$, the majoron equilibrates with the neutrinos. For the $1$ keV candidate shown, this occurs for couplings $\lambda \gtrsim 4 \times 10^{-12}$, corresponding to $\Gamma_{\rm eff} \gtrsim 1$. The right panel of \Fig{fig:Majoron_evolution} illustrates the evolution of the energy density in the majoron system for the same $1$ keV candidate. One can see both from the evolution of $\Delta N_{\rm eff}$ and from the evolution of the energy density that equilibrium is indeed attained for $\lambda \gtrsim 4 \times 10^{-12}$, as was expected from the simple comparison of the interaction rate. We have verified using the full solutions that the approximations adopted above are valid to high precision.

At the moment, it is not practical from a computational perspective to implement the evolution of the background for every sampled point in parameter space. To avoid this issue we derive fitting formulas to map the evolution of $\rho_{\nu\phi}$, valid for arbitrary values of $m_\phi$ and $\Gamma_{\rm eff}$. These equations have been implemented into {\tt CLASS} for a rapid evaluation of the background evolution. For the sake of reproducibility, we provide the fitting formula for the energy density of the neutrino-majoron system, expressed in terms of the majoron mass and $\Gamma_{\rm eff}$:
 \begin{align}
 \frac{\rho_{\nu\phi}}{T_\gamma^4} = \alpha_1(\Gamma_{\rm eff}) - \frac{10^{\alpha_2(\Gamma_{\rm eff})}}{1 + \alpha_3(\Gamma_{\rm eff}) \left(1 + \frac{T_\gamma}{m_\phi}\right) e^{-\left(\frac{T_\gamma}{m_\phi} - 1\right) / \alpha_4(\Gamma_{\rm eff}) } } \, ,
 \end{align}
 where $\Gamma_{\rm eff}$ is as defined in \Eq{eq:gamma_eff2} and with:
 \begin{align}
 \alpha_1(\Gamma_{\rm eff}) & =   0.473770 - \frac{0.0066356 \times  e^{\left(2.79816 \, \sqrt{\Gamma_{\rm eff}}  + 1\right) } }{1 + 3.25434 \, \sqrt{\Gamma_{\rm eff}} \, e^{\left(2.79816 \, \sqrt{\Gamma_{\rm eff}}+ 1\right) } } \, ,  \\
 \alpha_2(\Gamma_{\rm eff}) & =  -1.7329 - \frac{0.459096711 \times e^{\left(1.41748 \sqrt{\Gamma_{\rm eff}} + 1\right) }}{1 + 8.92332  \sqrt{\Gamma_{\rm eff}} e^{\left(1.41748 \sqrt{\Gamma_{\rm eff}}  + 1\right) } } \, , \\ 
 \log_{10}\left(\alpha_3(\Gamma_{\rm eff})\right) & =  -1.23587 \times {\Gamma_{\rm eff}}^{-0.2151} -  1.87406\times 10^{-6} \times \Gamma_{\rm eff}^{-1.52905} - 1.20918 \times \Gamma_{\rm eff}^{-0.2182} \, , \\
 \alpha_4(\Gamma_{\rm eff}) & =  -0.011777 + 0.15448\frac{\sqrt{\Gamma_{\rm eff}}}{1 + 0.738962 \sqrt{\Gamma_{\rm eff}}} + 0.04227\, e^{- \left(\frac{4 \sqrt{\Gamma_{\rm eff}} -1}{16.1314} \right)^2}  \, .
 \end{align}

\begin{figure}
\centering
\includegraphics[width=0.48\textwidth]{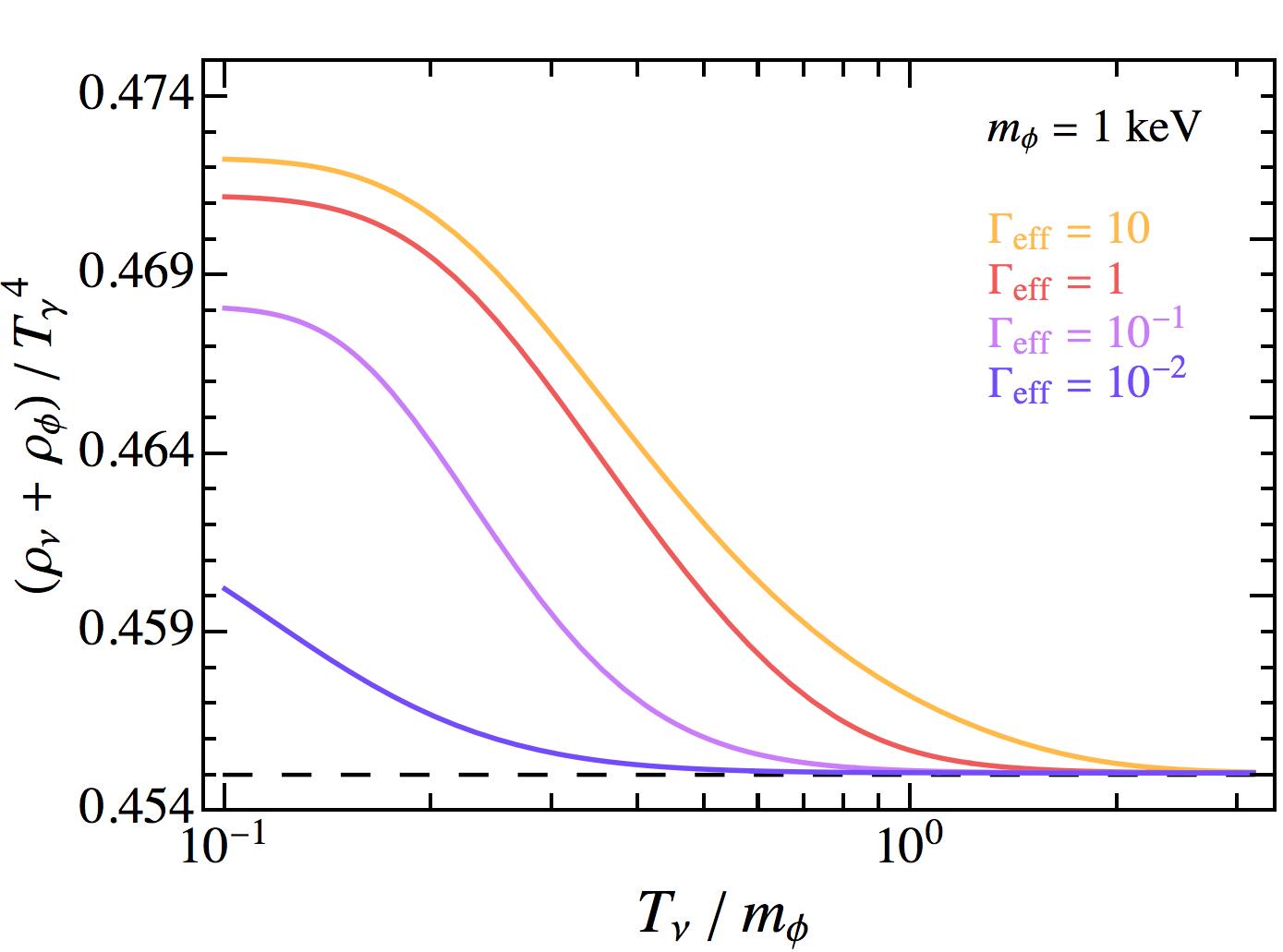}
\includegraphics[width=0.48\textwidth]{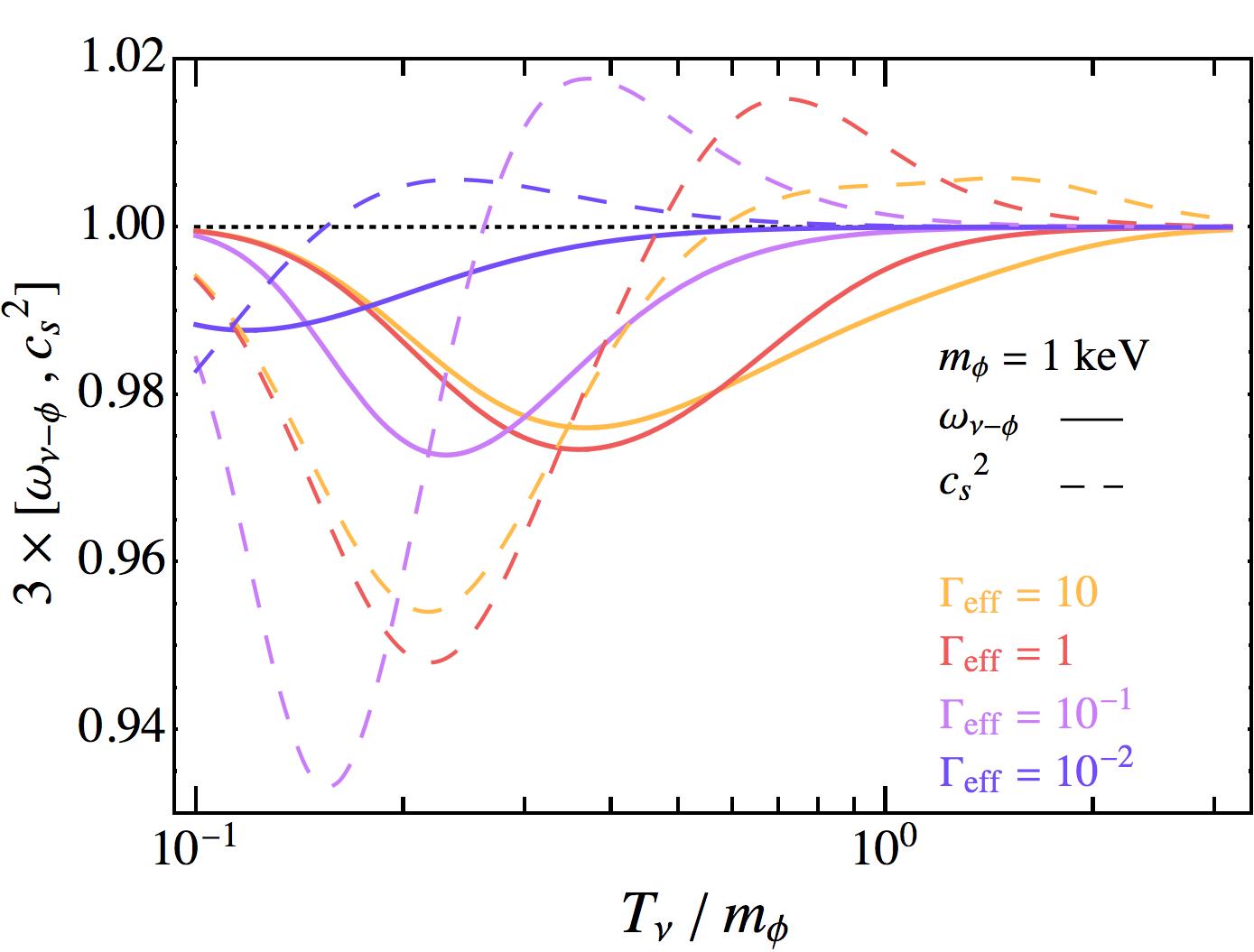}\vspace{-0.3cm}
\caption{\label{fig:EoS} Left: Evolution of the joint neutrino-majoron energy density. Right: Evolution of both the equation of state (solid) and speed of sound (dashed) for a majoron with $m_\phi = 1$ keV and various interaction strengths, with both normalized via a multiplicative factor of 3 such that for radiation $ 3 \times \omega = 3 \times c_s^2 = 1$. Black curves in both panels denote the $\Lambda$CDM values. }
\end{figure}

\begin{figure}
\includegraphics[width=0.48\textwidth]{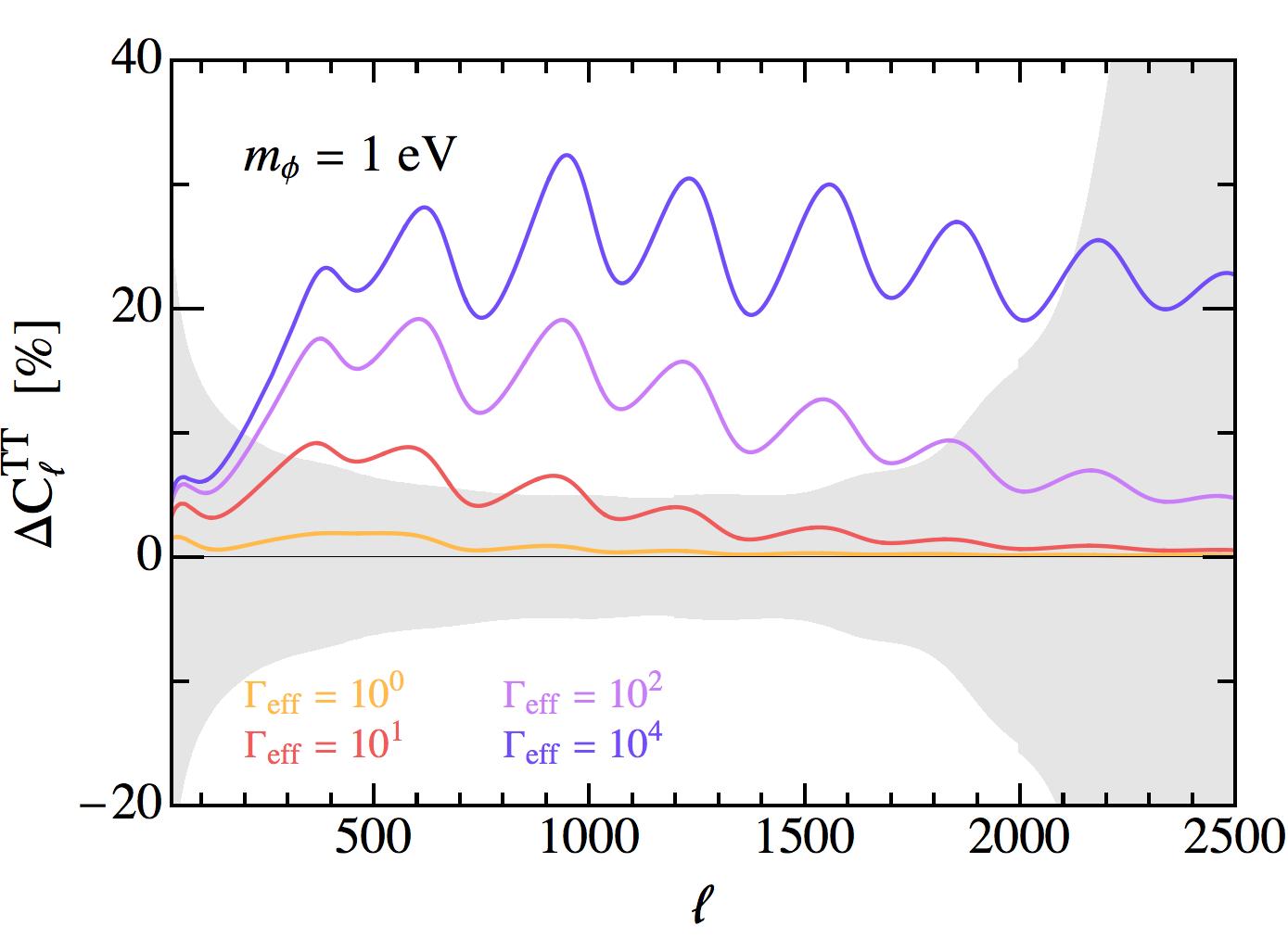}
\includegraphics[width=0.48\textwidth]{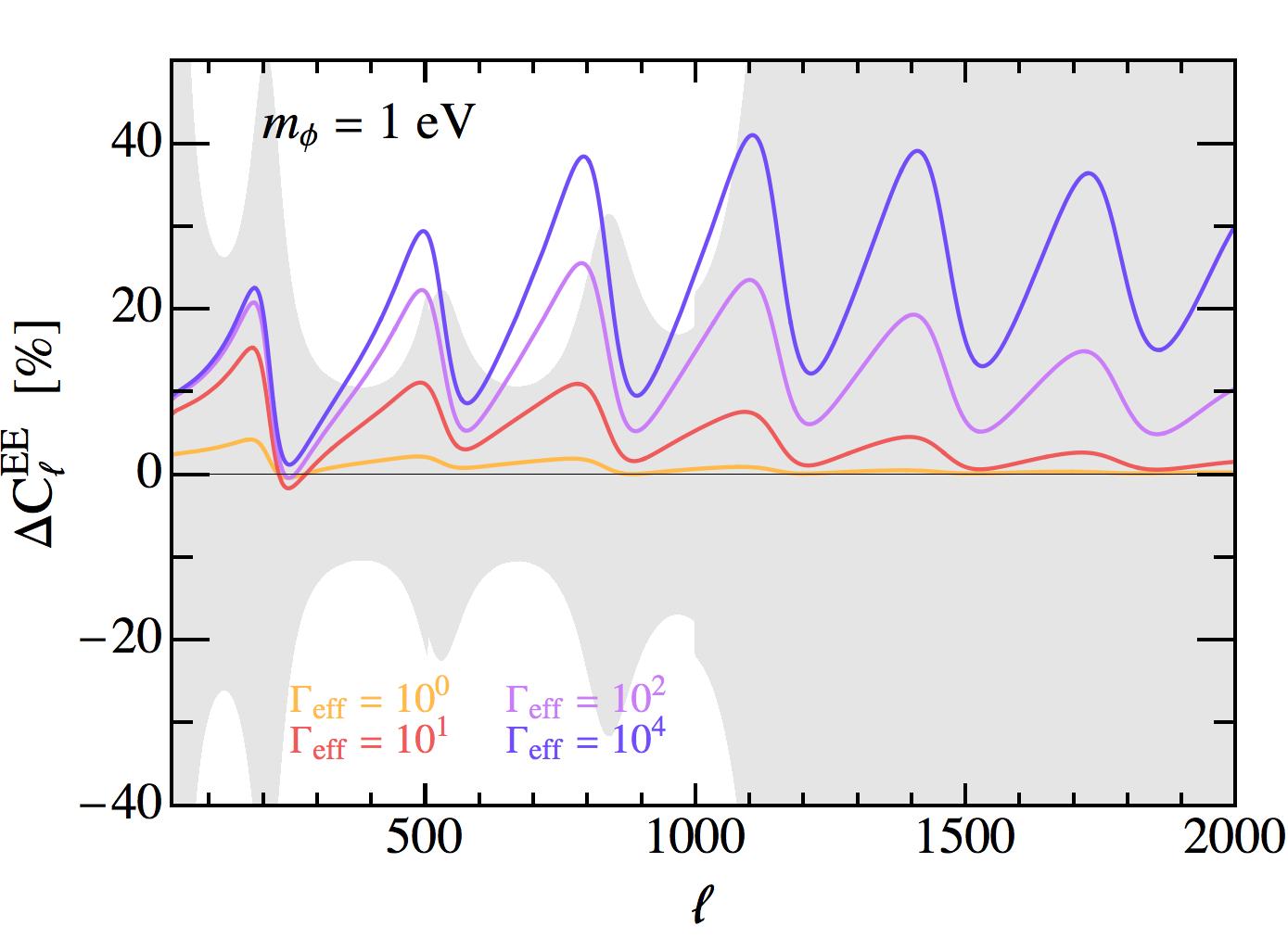}\vspace{-0.3cm}
\caption{\label{fig:cls_1ev} Percent difference between the TT (left) and EE (right) power spectrum for a $1$ eV majoron with various values of $\Gamma_{\rm eff}$. One sigma errors from Planck 2018 are shown in gray. }
\end{figure}
\begin{figure}
\includegraphics[width=0.48\textwidth]{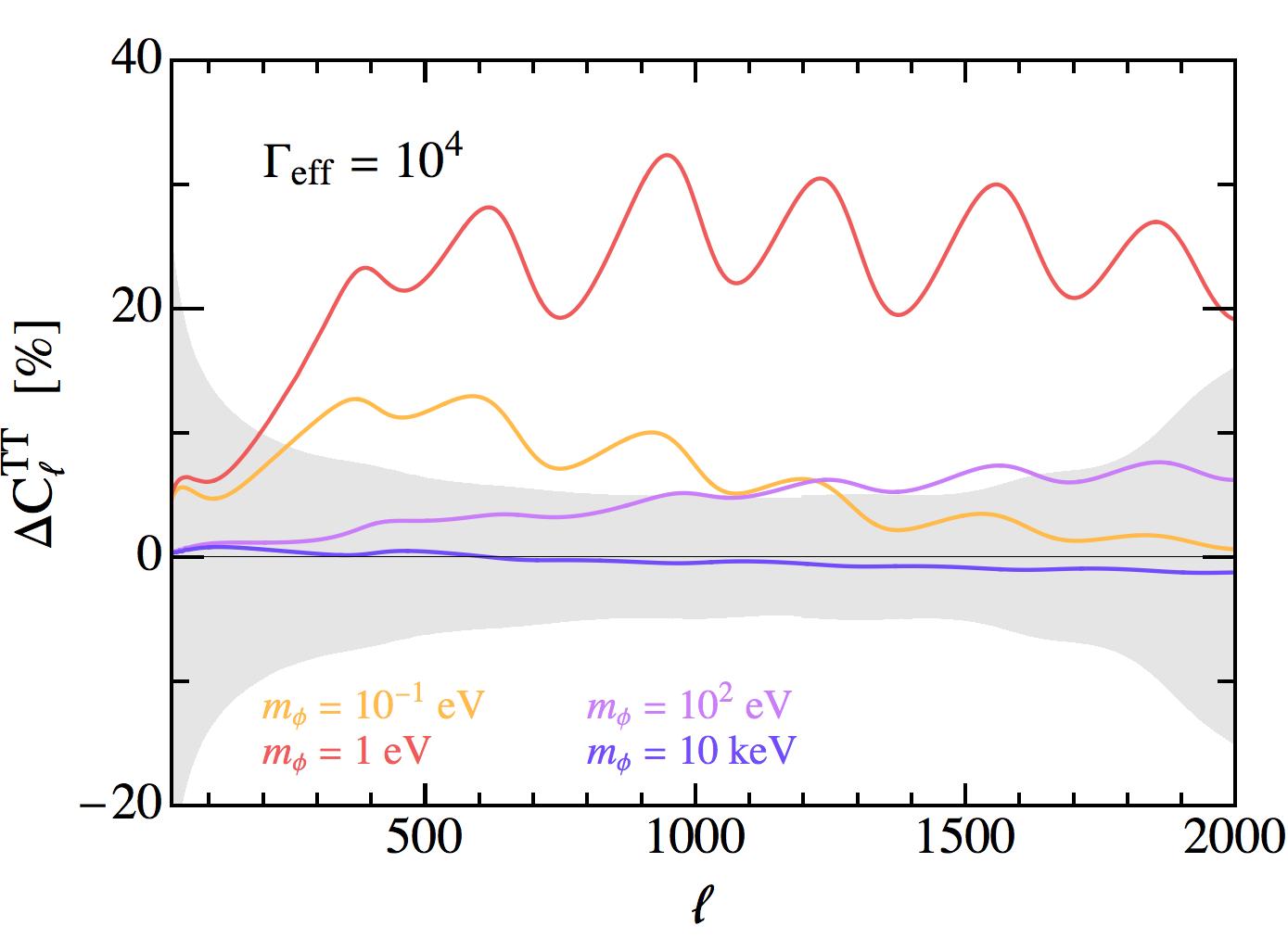}
\includegraphics[width=0.48\textwidth]{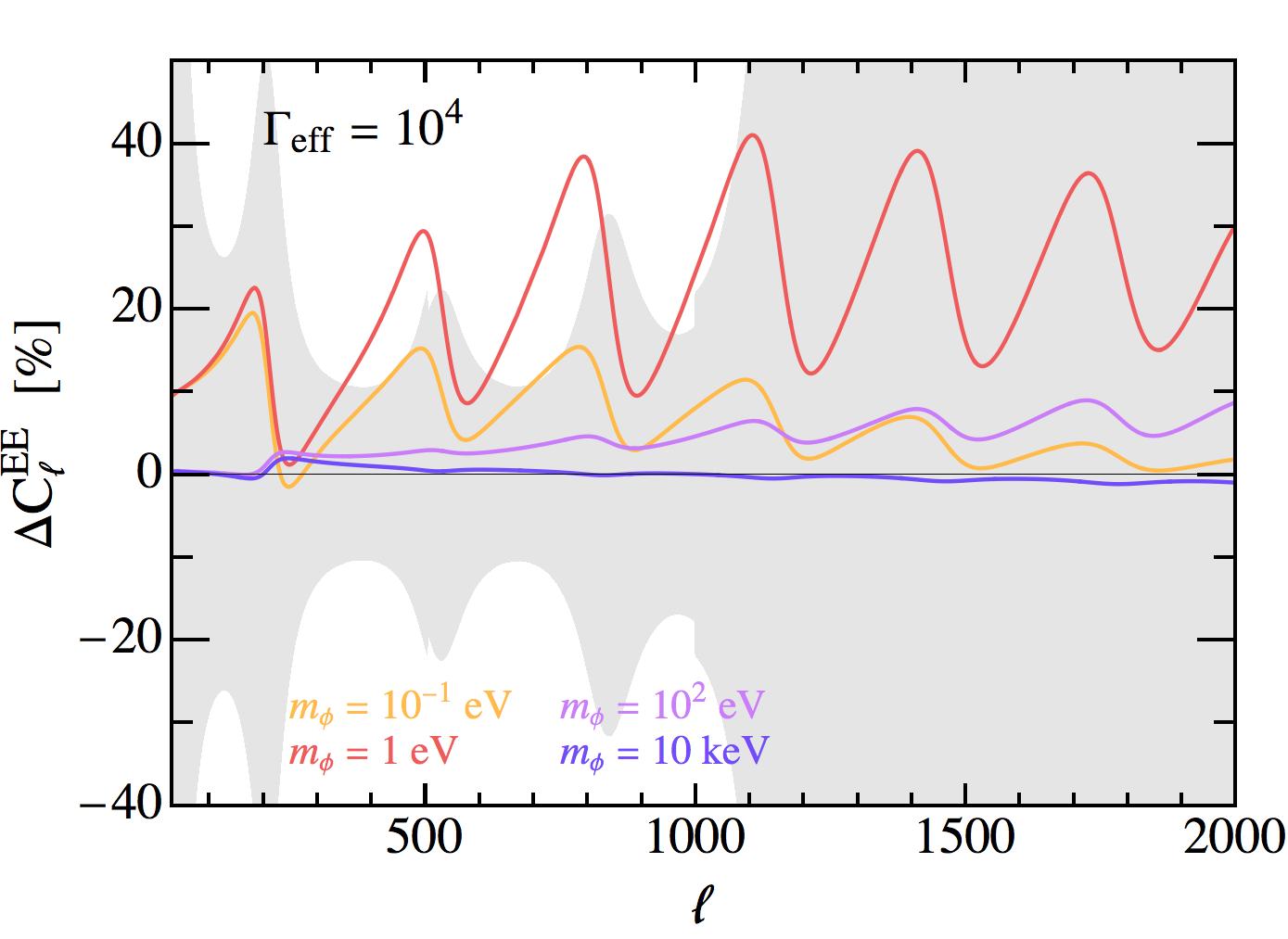}\vspace{-0.3cm}
\caption{\label{fig:cls_masses} Same as \Fig{fig:cls_1ev} but varying majorons masses $m_\phi$ from $10^{-1}$ eV to $10$ keV, keeping the effective interaction strength $\Gamma_{\rm eff} = 10^4$.  }\label{fig:mphivarying}
\end{figure}

\newpage 

~\\ \noindent{\bf CMB Phenomenology:}
For the numerical analysis presented in the main body of the text, we treat the neutrinos and majorons as a joint massless system. In order for this adopted treatment to be valid, the fractional shift in the energy density and equation of state from an ultra-relativistic system should be small. We illustrate in \Fig{fig:EoS} that indeed this approximation holds to extremely high degree, thus validating the joint treatment of these two species within a single massless fluid.

\begin{figure}
\includegraphics[width=0.48\textwidth]{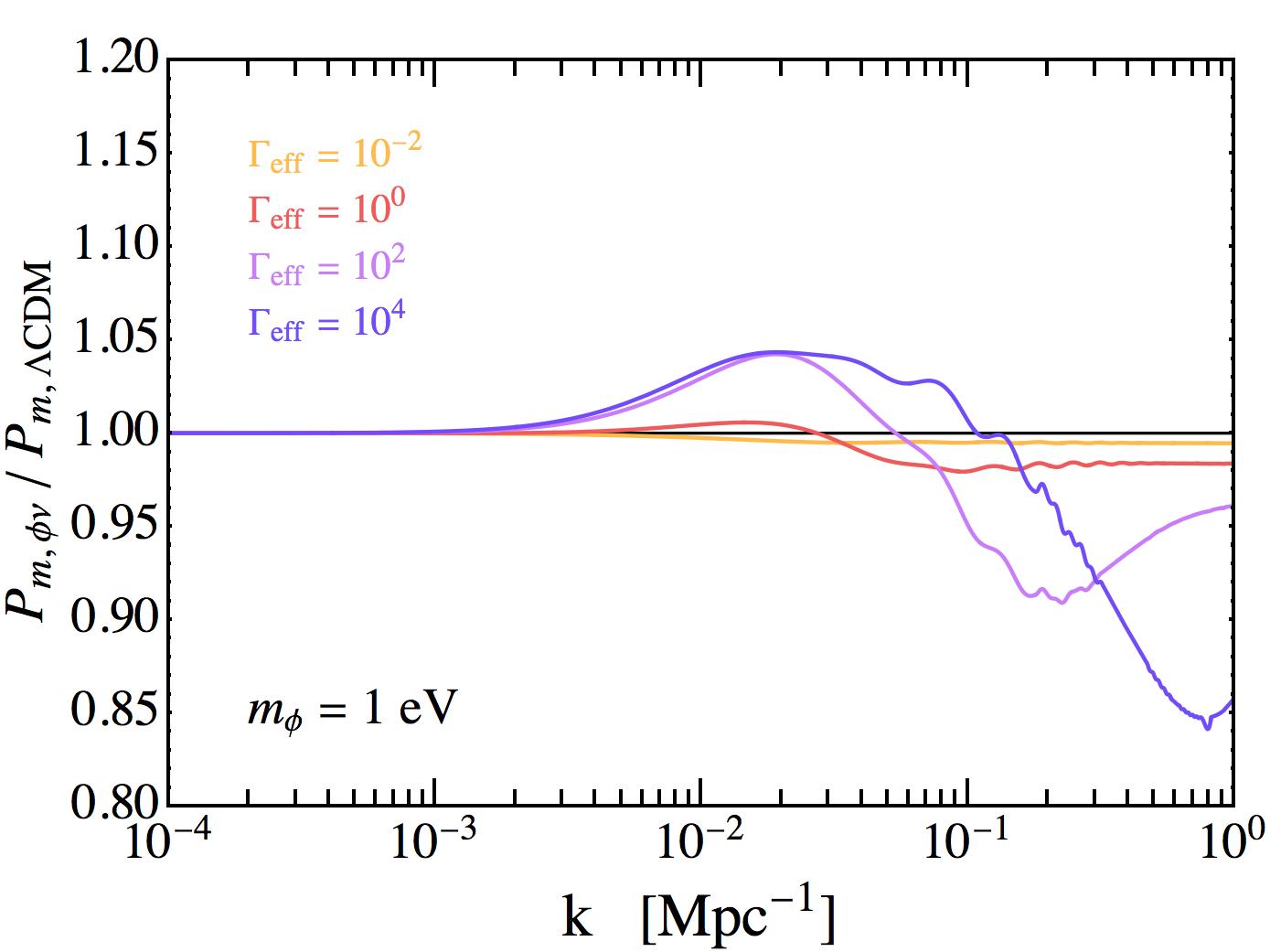}
\includegraphics[width=0.48\textwidth]{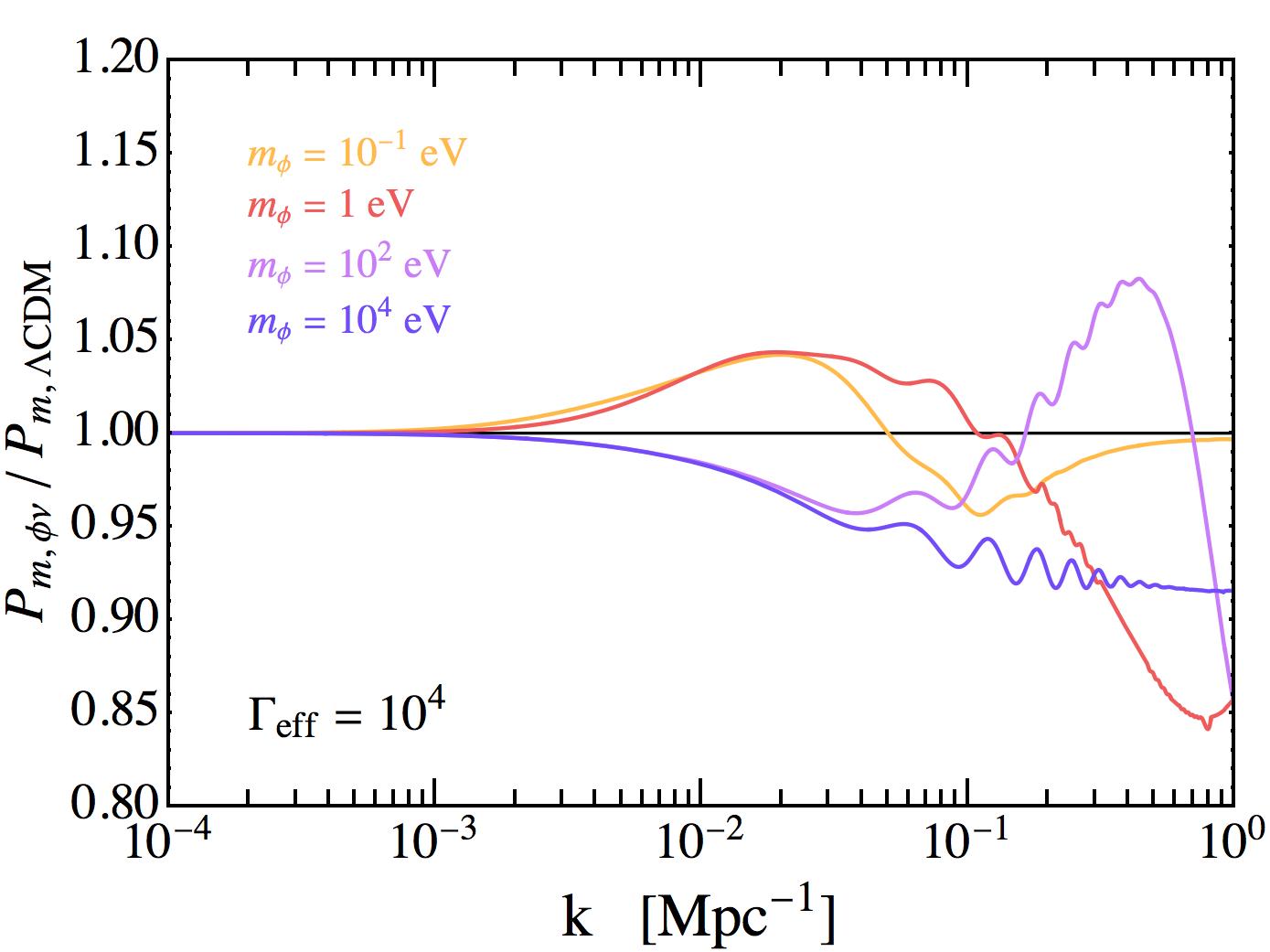}
\caption{\label{fig:mps} Ratio of the linear matter power spectrum in the interacting majoron scenario to that of $\Lambda$CDM for a $1$ eV majoron with various values of $\Gamma_{\rm eff}$ (left), and varying the majoron mass but keeping $\Gamma_{\rm eff} = 10^4$ (right), normalized at large scales. }
\end{figure}

In Figs.~\ref{fig:cls_1ev} and \ref{fig:cls_masses}, we illustrate the impact of the majoron on the TT and EE power spectra for $m_\phi= 1\,\text{eV}$ and various values of $\Gamma_{\rm eff}$ (\Fig{fig:cls_1ev}), and $\Gamma_{\rm eff} = 10^4$ with various values of $m_\phi$ (\Fig{fig:cls_masses}). For sufficiently light majorons, presence of interactions enhances both the TT and EE spectra, and induces periodic oscillations in the $\mathcal{C}_\ell$'s. For large masses, the impact of the perturbations vanish and the remaining signature is simply that induced by the presence of an additional contribution to $\Delta N_{\rm eff}$. For completeness, we also show in \Fig{fig:mps} the relative change in the linear matter power spectrum induced at small scales for the same candidates.

\begin{figure}
\centering
 \includegraphics[width=0.48\textwidth]{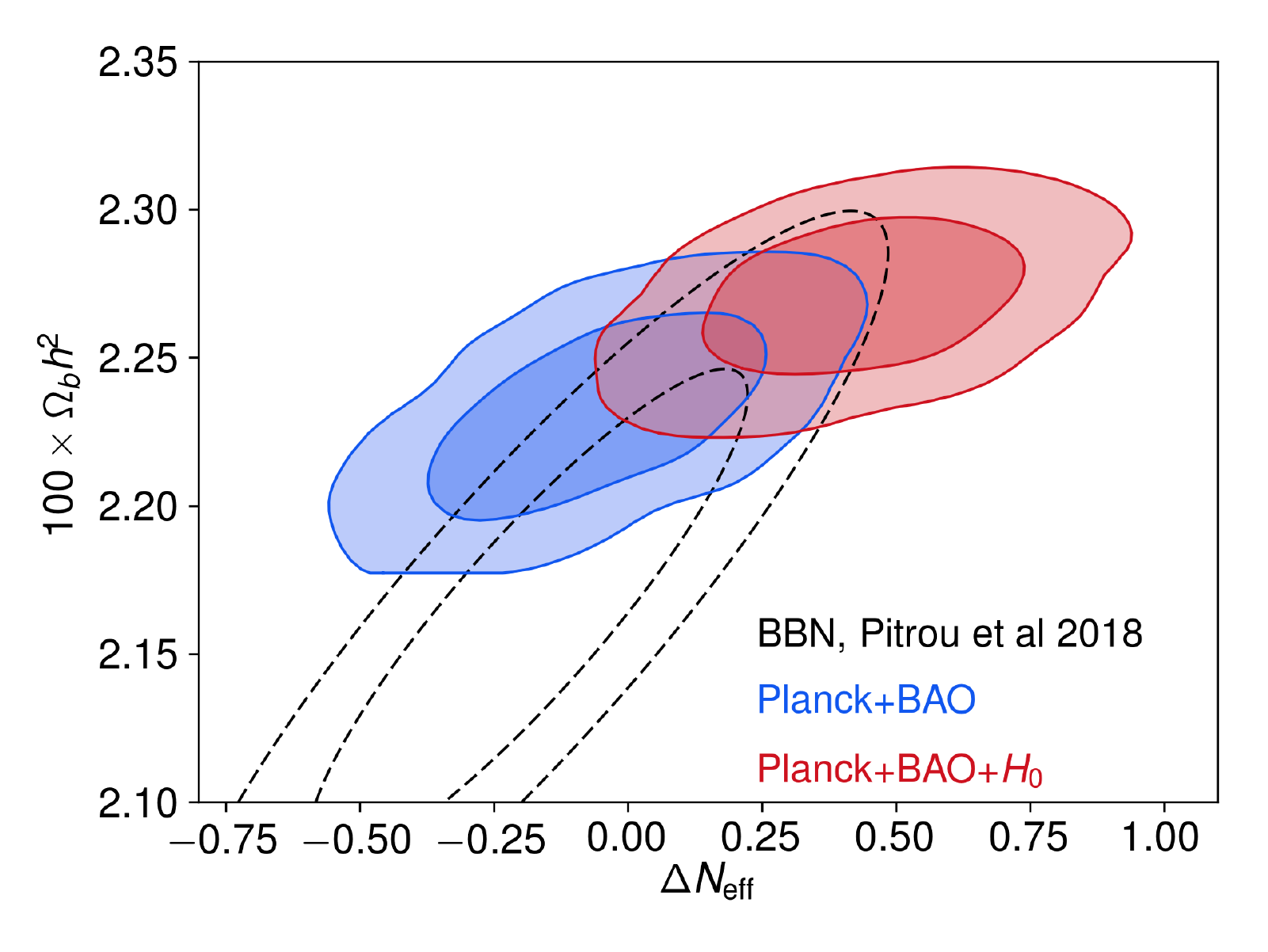}\vspace{-0.4cm}
\caption{1$\sigma$ and 2$\sigma$ contours for $\Omega_bh^2$ and $\Delta N_{\rm eff}$ using the measured primordial element abudances~\cite{Pitrou:2018cgg} (black dashed), and compared with the preferred regions in a $\Lambda$CDM+$\Delta N_{\rm eff}$ scenario obtained using the Planck+BAO and Planck+BAO+$H_0$ likelihood analysis.}\label{fig:omegab-vs-Neff}
\end{figure}

As discussed in the primary text, including a floating value of $\Delta N_{\rm eff}$ shifts the preferred value of $\Omega_b$ which is probed by the CMB and BBN. Obtaining a coherent cosmological picture requires ensuring compatibility of these two distinct probes with local measurements of $H_0$. $N_{\rm eff}$ can be modified after BBN and prior to recombination, as \eg is done in the case of the majoron. Larger values of $N_{\rm eff}$, as preferred to resolve the $H_0$ tension, naturally shift $\Omega_b$ to larger values, however the degeneracy of these parameters in the CMB and BBN is not exact for the case of $\Lambda$CDM + $\Delta N_{\rm eff}$. This is shown explicitly in \Fig{fig:omegab-vs-Neff}, and we note that the case of the majoron + $\Delta N_{\rm eff}$ is quite similar to the case of the$\Lambda$CDM + $\Delta N_{\rm eff}$. It is interesting that as the value of $H_0$ shifts toward the locally measured value (as occurs when one includes the SH$_0$ES dataset in the likelihood), the preferred central value derived from the CMB analysis produces an increasing tension with the values inferred from BBN. While this tension is mild, it is important to bare in mind that the central value of $H_0$ in the {\tt Planck}+BAO+SH$_0$ES analysis is still reasonably below the central value preferred by the SH$_0$ES data itself.

\begin{figure}[t]
\centering
\hspace{-0.8cm} \includegraphics[width=0.9\textwidth]{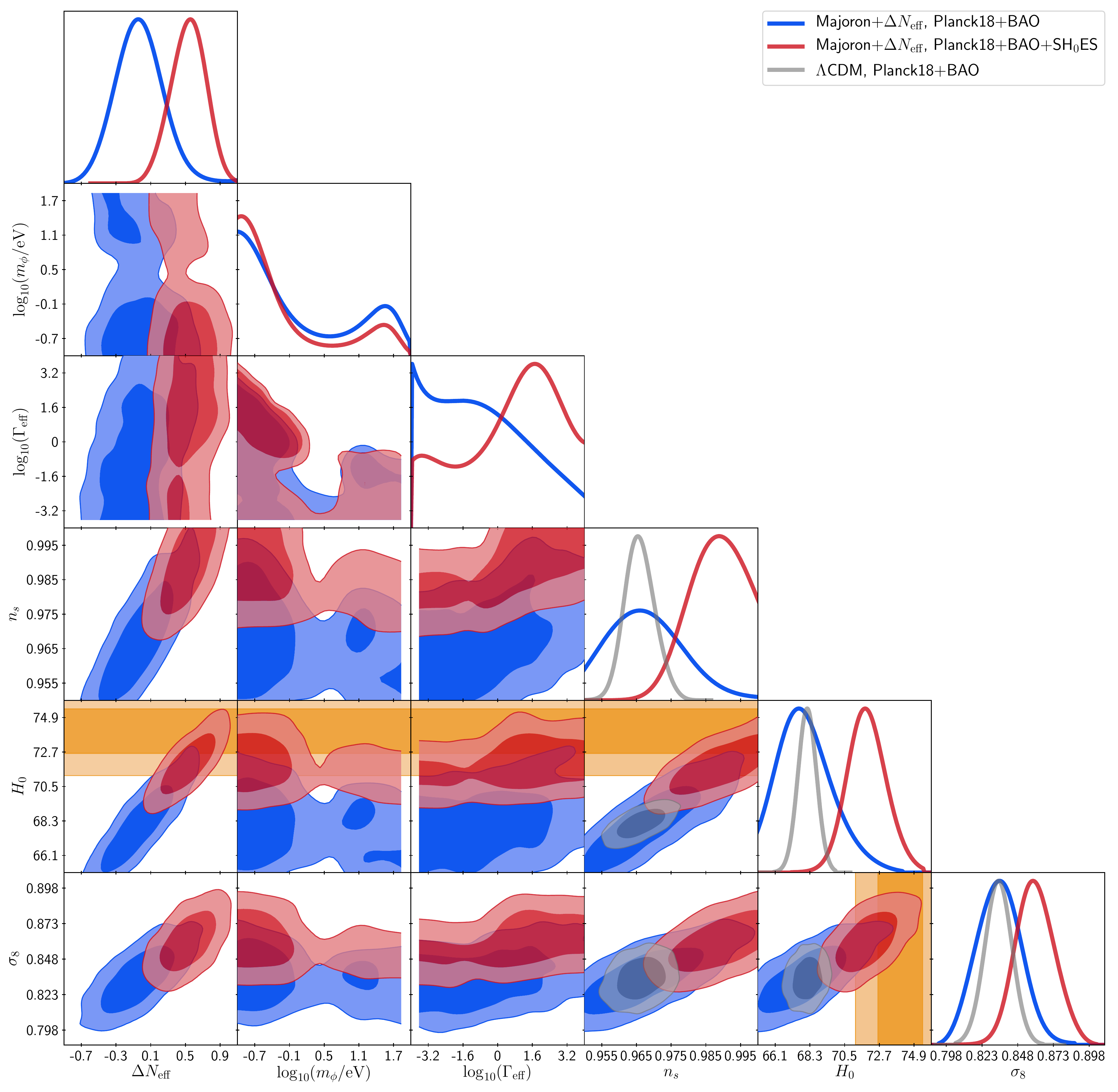}  \vspace{-0.3cm}
\caption{Corner plot displaying $1\sigma$ and $2\sigma$ two-dimensional posteriors for $\Lambda$CDM (grey), and the majoron + $\Delta N_{\rm eff}$ model, including (red) and excluding (blue) the SH$_0$ES likelihood. Shown for comparison are the 1 and 2$\sigma$ SH$_0$ES posterior (orange).}\label{fig:2D}
\end{figure}

Finally, we show the two-dimensional posterior corner plot in \Fig{fig:2D}. In the $\Gamma_{\rm eff}$ vs $m_\phi$ two-dimensional posterior a double peak structure in the majoron mass can be seen. This is a result of the fact that majoron-neutrino perturbations particularly affect the CMB spectra when $m_\phi \sim 1 \,\text{eV}$ (as can be appreciated from Figure~\ref{fig:mphivarying} and from the posterior). This leads to stringent constraints on $\Gamma_{\rm eff}$ for $m_\phi \sim 2 \,\text{eV}$ as highlighted in Figure~\ref{fig:Majoron_easy}, and to a double peak posterior on $m_\phi$ with maximums at $m_\phi \sim 0.3\,\text{eV}$ and $m_\phi \sim 30\,\text{eV}$.

~\\ \noindent{\bf Big Bang Nucleosynthesis Constraints:}
We set constraints on the neutrino-majoron coupling by requiring successful BBN by using the constraint on the effective number of relativistic degrees of freedom during BBN. The latest analysis finds~\cite{Pitrou:2018cgg}:
\begin{align}
	N_{\rm eff}^{\rm BBN} = 2.88 \pm 0.27 \,\qquad 68\% \,\text{CL} \, .
\end{align}
This means that the one-sided 95\% CL upper limit is $N_{\rm eff}^{\rm BBN} < 3.33$. Since we do not explicitly solve for the light element abundances within our modified cosmology, we conservatively adopt an upper limit of $N_{\rm eff}^{\rm BBN} < 3.5$. It is worth noting that majorons in thermal contact with neutrinos right before the time of neutrino decoupling would induce a shift in $N_{\rm eff}$ at the level of $\Delta N_{\rm eff} = N_{\rm eff}-N_{\rm eff}^{\rm SM} = 4/7 \simeq 0.57$ (where $N_{\rm eff}^{\rm SM} = 3.045$~\cite{Escudero:2020dfa,deSalas:2016ztq,Mangano:2005cc}), which is clearly excluded by the measured primordial element abundances. 

The main effect of a positive contribution to $\Delta N_{\rm eff}$ at the time of BBN is to induce a higher expansion rate of the Universe during the formation of the primordial elements with respect to the SM. Thus, the bound from BBN on $\Delta N_{\rm eff}$ can be interpreted as a time constraint on the generation of the primordial element abundances. In particular, in a Universe with $\Delta N_{\rm eff} = 0.45$, the time at which Deuterium forms (corresponding to $T_\gamma^{\rm D} \simeq 0.07\,\text{MeV}$~\cite{Pospelov:2010hj,Iocco:2008va,Sarkar:1995dd}) is $t = 256.69\,\text{s}$\footnote{In the SM, $t_{T_\gamma = 0.07\,\text{MeV}} = 264.60\,\text{s}$.}. Consequently, we derive the BBN constraint shown in \Fig{fig:Majoron_easy} by requiring
\begin{align}\label{eq:bbn_C}
	T_\gamma > 0.07\,\text{MeV} \,,\qquad \text{at} \qquad t = 256.69\,\text{s} \, .
\end{align}

Any appreciable change in the expansion history of the Universe for $T_\gamma \lesssim T_\gamma^{\rm D} $ has been shown to render a minor impact on any relevant primordial nuclei abundance~\cite{Berlin:2019pbq}. We find that imposing the constraint in \Eq{eq:bbn_C} leads to the following bound on the majoron-neutrino coupling strength:
\begin{align}
	\lambda < \frac{1}{(5\times 10^{-9} \,\frac{\text{MeV}}{m_\phi})^{-1} + \left(4\times 10^{-5}\right)^{-1}}\,,
\end{align}
where the first term in the denominator results from the majoron production by inverse neutrino decays and the second term (\ie $\lambda < 4\times 10^{-5}$) arises from the production of majorons via neutrino-neutrino annihilations (see~\cite{Escudero:2019gfk}).

~\\ \noindent{\bf Primordial Majoron Abundance:}
Majoron interactions with the Standard Model arise through the active-sterile neutrino mixing~\cite{Chikashige:1980ui}, implying that majorons have non-negligible interactions with heavy sterile neutrinos. If the Universe was reheated to sufficiently high temperatures, it is conceivable that a primordial thermal population of majorons can be produced as a result of these interactions~\cite{Akhmedov:1992hi}. Here we comment under which conditions this occurs and the cosmological implications of such a primordial majoron population.

Within the singlet majoron scenario, sterile neutrinos decay into an active neutrino and a majoron at a rate~\cite{GonzalezGarcia:1990fb,Pilaftsis:1991ug}:
\begin{align}
\frac{\Gamma(N\to \phi \nu)}{\Gamma(N \to {\rm SM})}  \sim \left(\frac{v_L}{v_H}\right)^2 \qquad \text{if} \qquad m_N > m_W, \, \qquad \qquad \frac{\Gamma(N\to \phi \nu)}{\Gamma(N \to {\rm SM})} \sim \left(\frac{v_L}{v_H}\right)^4 \qquad \text{if} \qquad m_N < m_W \,.
\end{align} 
Which implies that sterile neutrinos will have sizable decays to majorons provided that $v_L > v_H$, or equivalently $\lambda \lesssim 10^{-13}$. 

The production rate of sterile neutrinos from the SM plasma is $\Gamma \sim 4 \times 10^{-3} \, y_N^2 \, T $~\cite{Besak:2012qm,Garbrecht:2013urw,Ghisoiu:2014ena}, where $y_N$ is the sterile neutrino Dirac Yukawa coupling -- which, within the type-I seesaw is $y_N \sim 4\times 10^{-8} \sqrt{m_N/\text{GeV}}\sqrt{m_\nu/0.05\,\text{eV}}$. By comparing the Hubble parameter $H \sim 1.66 \sqrt{g_\star} \, T^2/M_{\rm pl}$ with the sterile neutrino production rate, it is easy to show that sterile neutrinos (with couplings capable of generating the observed neutrino masses) are brought into thermal equilibrium at temperatures
\begin{align}
T \sim 5 \times m_N  \times \left({m_\nu}/{0.05\,\text{eV}}\right) , 
\end{align} 
and would disappear from the plasma soon after they become non-relativistic, at $T \sim m_N/3$. Clearly, if such sterile neutrinos decay into majorons, they will produce a primordial thermal population of these particles. This statement is, however, dependent upon the unknown thermal history of the Universe -- for example, this can be trivially avoided if the reheating temperature $T_{\rm RH} < m_N$, as it would prevent sterile neutrinos from ever being thermalized in the early Universe. 

Once sterile neutrinos decay/annihilate away from the thermal plasma, the majoron bath decouples from the SM model plasma. The majoron temperature after electron-positron annihilation is simply given by entropy conservation and reads:
\begin{align}
\frac{T_\gamma}{T_{\phi}} \simeq \frac{1}{3}  \left(\frac{{g_{\star S}^{\rm SM}|_{\rm today}}}{3.93}\right)^{1/3} \,  \left(\frac{106.75}{g_{\star S}^{\rm SM}|_{\phi-\text{decoupling}}}\right)^{1/3}  \,,
\end{align}  
which corresponds to $\Delta N_{\rm eff} = 0.027$ at the time of BBN, provided $m_\phi < 1\,\text{MeV}$. 

\begin{figure}
\centering
\hspace{-0.8cm} \includegraphics[width=0.5\textwidth]{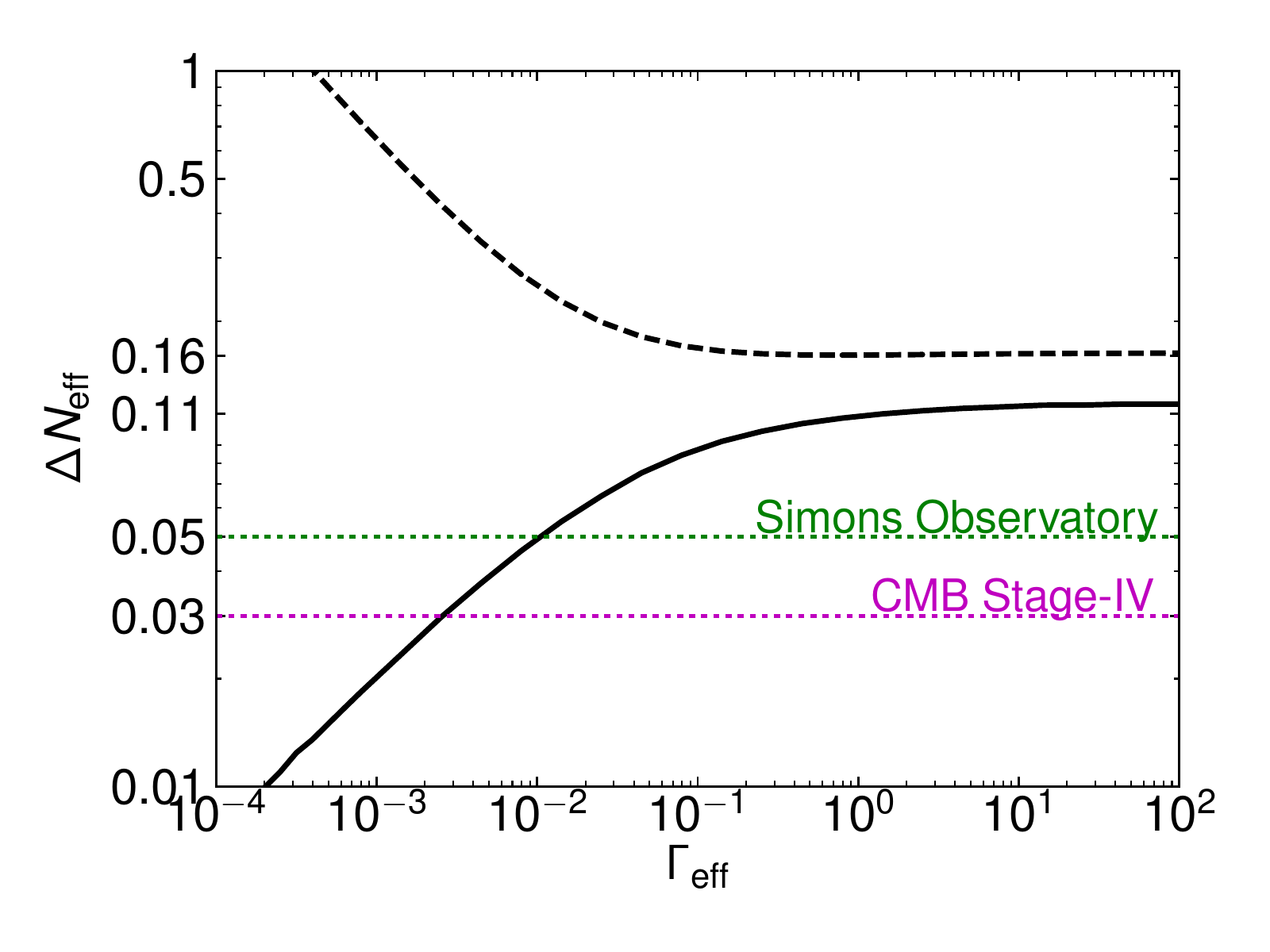}  \vspace{-0.4cm}
\caption{$\Delta N_{\rm eff}$ as a function of the effective neutrino-majoron interaction rate with $m_\phi = 1\,\text{keV}$. The dashed line corresponds to the scenario in which a thermal population of majorons was produced in the very early Universe. The solid line corresponds to the scenario in which there is no primordial population of majorons and they are solely produced by neutrino inverse decays. Sensitivities are shown for Simons Observatory~\cite{Ade:2018sbj} and CMB-S4~\cite{Abazajian:2016yjj} at the $1\sigma$ level.  }\label{fig:Neff_Thermal}
\end{figure}

In \Fig{fig:Neff_Thermal} we compare the relative contribution to $N_{\rm eff}$ as a function of $\Gamma_{\rm eff}$, assuming {\emph{(i)}} a small pre-existing thermal population of majorons present at early times (that decoupled at $T \gtrsim 100\,\text{GeV}$) and {\emph{(ii)}} majorons are only produced via inverse decays of neutrinos. Projected sensitivity from the Simons Observatory~\cite{Ade:2018sbj} and CMB-S4~\cite{Abazajian:2016yjj} are shown for comparison. For small majoron interactions, a pre-existing thermal population comes to dominate the energy density and produces a large shift in $\Delta N_{\rm eff}$ that can be easily constrained by observations of the CMB. 

In \Fig{fig:Majoron_easy}, we include a black dotted line that denotes the region of parameter space for which a primordial majoron population would produce $\Delta N_{\rm eff} \geq 1$ at recombination, and would thus be excluded by {\tt Planck}. To be concrete, we determine the bound at:
\begin{align}
\Gamma_{\rm eff} > 4\times 10^{-4}\,,\qquad \lambda > 8\times 10^{-14} \sqrt{\frac{m_\phi}{\text{keV}}} \,, \qquad v_L < 1.2 \,\text{TeV} \, \sqrt{\frac{\text{keV}}{m_\phi}} \, .
\end{align}

Finally, notice that presence of a primordial population of majorons would lead to enhanced damping to the neutrino anisotropic stress and an additional contribution to $\Delta N_{\rm eff}$ (such that if $\Gamma_{\rm eff} > 1$, $\Delta N_{\rm eff}$ would asymptote to $0.16$ rather than $0.11$) -- in this scenario, we expect that the constraints derived in this work to strengthen. 

\end{document}